\documentstyle[11pt,twoside]{article}
\begin{document}\hbadness=10000\thispagestyle{empty}
\pagestyle{myheadings}
\markboth{H.-Th. Elze, T. Kodama and R. Opher}
{Collective Modes in Neutrino `Beam' Electron-Positron Plasma Interactions}
\title{{\bf Collective Modes in Neutrino `Beam' Electron-Positron Plasma Interactions}}
\author{$\ $\\
{\bf Hans-Thomas Elze$^{1}$, Takeshi Kodama$^1$ and Reuven Opher$^2$}\\ $\ $\\ 
$^1$Universidade Federal do Rio de Janeiro, Instituto de
F\'{\i}sica\\
\,Caixa Postal 68.528, 21945-970 Rio de Janeiro, RJ, Brazil \\
\rule{0cm}{0.7cm} and \\ \rule{0cm}{0.6cm} 
$^2$Instituto Astron\^omico e Geof\'\i sico, USP\\
\,\,\,\,\,\,Caixa Postal 9638, 01065-970, S\~ao Paulo, SP, Brazil
}
\vskip 0.5cm
\date{June 2000}
\maketitle 
\vspace{-8.5cm}
\vspace*{8.0cm}
\begin{abstract}{\noindent 
We derive semiclassical neutrino-electron transport equations in the collisionless (Vlasov) limit from the 
coupled Dirac equations, incorporating the charged and neutral weak current-current as well as electromagnetic 
interactions. A corresponding linear response theory is derived. In particular, we calculate the response 
functions for a variety of beam-plasma geometries, which are of interest in a supernova scenario. 
We apply this to the study of plasmons and to a new class of collective 
{\it pharon} resonance modes, which are characterized by $\omega < q$.   
We find that the growth rates of the unstable modes correspond to a 
strongly temperature ($\propto T_\nu^2T_e^3$) and linearly momentum dependent e-folding length of 
about $10^{10}$\,km under typical conditions for Type\,II supernovae. This appears to rule out such 
long-wavelength collective modes as an efficient means of depositing neutrino energy into the plasma sphere.
}\end{abstract}
\section{Introduction} 
Neutrino transport processes are known to play a major role in the 
energy-momentum flow powering the dynamics of Type\,II supernovae \cite{Cooperstein88,RECENT1,RECENT2}. Generally, it has been that the collision-dominated aspects have been studied in detail, 
leading to substantial progress in the understanding of these stellar explosions, while, however, still leaving open some problems in the quantitative description of their spatio-temporal (hydrodynamic) evolution.  
  
An earlier work by Bethe \cite{Bethe86}, which introduced the idea of a modified in-medium neutrino dispersion relation and of a corresponding  effective Hamiltonian, was used in the series of papers \cite{Bingham94} to describe the collective interaction of an intense neutrino flux (from the supernova core) with an electron-positron plasma (the supernova atmosphere) of comparatively low temperature. Their tentative conclusion was that a particular induced plasma instability may be much more efficient than traditional collision dominated mechanisms, i.e., faster by many orders of magnitude, in depositing the neutrino energy into the plasma sphere \cite{Bingham94}.
  
However, the approach in these papers is subject to criticism, since there is no physical or formal justification (such as a hypothetical condensate) for the scalar `bosonic' collective neutrino wave function used.
In particular, the implied quantum phase coherence of the neutrinos appears hard to justify. Considering their incoherent thermal production and the effective duration or length of the `beam', no bunching effects are to be expected. Moreover, it is somewhat hidden in their phaenomenological approach how the precise V-A tensor  
structure of the electroweak current-current interactions can be taken into account.   
This has also been pointed out in Ref.\,\cite{Bento99} recently. Employing finite temperature field theory, Bento studies the excitation/damping of longitudinal electromagnetic plasmons (Langmuir modes) in the electron plasma under the influence of a neutrino flux. His results indicate that this type of collective mode instabilities ``... do not seem to be a viable mechanism of substantial energy transfer from neutrinos to a supernova plasma'' \cite{Bento99}. We confirm this.         
  
It is the purpose of our present work to systematically derive the transport equations for the neutrino-electron system from first principles (Section 2), as well as the relevant dispersion relations (Section 3). We introduce appropriate 
spinor Wigner functions, while deriving the detailed chiral structure of the neutrino Wigner 
function in the Appendix. 
Previously, only the phenomenological approach \cite{Bingham94} or the perturbative 
finite-temperature field theory \cite{Bento99,Nieves00} were applied. Our 
general derivations may also prove useful for other astrophysical applications, such as those involving strong magnetic fields or, generally, neutrino transport under mean field conditions.

In the collisionless regime, the results of Sections 2 and 3 allow us to investigate, in detail, the collective modes in the highly anisotropic neutrino `beam' plus electron-positron plasma system. We find longitudinal and transverse plasmons, which are only perturbatively modified by the neutrino flux. 

Furthermore, we also find a new class of growing, as well as decaying collective oscillations, nonexistent in isotropic equilibrium plasmas, which we name `{\it pharons}'\footnote{After the island of {\it Pharos}, where the famous lighthouse of ancient Alexandria was constructed under the order of Ptolemeus II.}. They are caused by a resonance effect, generally at a frequency $\omega$ less than the momentum $q$, due to the unbalanced neutrino momentum distribution, which is characterized by a finite opening angle with respect to the beam axis. We study such modes with the wave vector parallel to the beam direction (Type\,I pharons), as well as with the wave vector orthogonal to the beam direction (Type\,II pharons). 
    
Geometrically, the Type\,II situation corresponds most closely to the one where the two-stream instabilities would be expected to occur in ordinary plasmas. We investigate whether such instabilities are induced by the weak current-current interactions and, depending on the growth rates, may provide an essential contribution to the still partly elusive energy transfer mechanism in Type\,II supernovae (Section 3). 

The collective two-stream filamentation instability is well known to occur in ordinary plasmas due to the electromagnetic Lorentz force \cite{Weibel59,LifshitzP81}. More recently, it has also been studied in the context of strong (color-electromagnetic) interactions, where two interpenetrating parton beams describe high-energy nuclear reactions, see \cite{Stan96} and further references therein. 

More generally, one may expect such `hydrodynamic' instabilities in interacting many-body systems, in particular, in plasmas with interactions 
mediated by the Standard Model gauge fields, whenever the system consists 
of two or more 
components with considerably different momentum space distributions \cite{Weibel59,LifshitzP81,Stan96}. In these cases, perturbations which, loosely speaking, are transverse to a predominant collective flow, tend to be amplified by the collective feedback effect of the effective long-range forces of mean field type.  

Consequently, we are motivated in this study by the supernova geometry, where the radially outward streaming neutrinos interact with the 
electron-positron plasma,  
which may produce a variety of collective instabilities.

\section{The Coupled Neutrino-Electron Transport Equations}
Our derivation of mean field transport equations will follow the 
successful strategy developed earlier for QED \cite{VGE87}, QCD \cite{EGV86}, and hadronic matter \cite{EGVetal87}. The basic idea 
can be easily summarized as follows: Starting from the underlying field 
operator equations of the model under consideration, one converts these 
into corresponding Wigner operator equations, i.e. for the density 
operator in the Wigner representation. In the appropriate 
mean field approximation the latter can be converted into a closed 
set of Wigner function equations (cf. Section 2.1); furthermore,  
performing a consistent $\hbar$-expansion, the most 
relevant semiclassical (Vlasov type) transport equations for 
the coupled relativistic phase space distributions  
are obtained (cf. Section 2.2). -- Presently, our notation and conventions follow those 
of Ref.\,\cite{VGE87}.

\subsection{From Dirac's to Mean Field Quantum Transport Equations}
We add the effective local coupling terms, 
\begin{equation}\label{Lint}
{\cal L}_{\mbox{int}}\equiv -\frac{G_F}{\sqrt 2}\left [
\bar\psi^{(\nu )}\gamma_\mu (1-\gamma_5)\psi^{(\nu )}\right ]
\left [\bar\psi^{(e)} \gamma^\mu (c_V-c_A\gamma_5)\psi^{(e)} 
\right ] 
\;\;, \end{equation} 
to the free Lagrangian densities of the electrons and electron 
neutrinos (including their antiparticles). 
They represent the {\it weak charged and neutral current-current interactions}   
of the Standard Model in the appropriate low-energy limit, with $c_V=\frac{1}{2}+2\sin^2\theta_W$ and $c_A=\frac{1}{2}$, using standard notation \cite{Bento99}. For $\mu$- and $\tau$-neutrinos, with only 
the neutral current interaction contributing, 
$c_{V,A}\rightarrow c_{V,A}-1$.
Here the neutrinos are described by four-component spinors as well and we allow   
for a small but finite neutrino mass, taking the growing evidence into account \cite{Roulet99}. 
Eventually, however, we will pass to the massless limit, since for our applications the masses of order eV are 
definitely negligible compared to the relevant energy scales of order MeV.  
  
The electromagnetic interaction of the electrons must 
be added to the interaction of Eq.\,(\ref{Lint}). For the derivation 
of the transport equations, however, this does not introduce a new 
element. The corresponding modifications will be added  
at the end of this section, making use of the earlier QED 
results \cite{VGE87}.  
 
The resulting Dirac (Heisenberg operator) equations for the 
electrons and neutrinos, 
incorporating the interaction (\ref{Lint}) in the mean field (Hartree) 
approximation, can be written in the form: 
\begin{equation}\label{Dirac} 
\left\{ i\gamma\cdot\partial_x-m^{(l)}-J^{(l')}\cdot\gamma
(c_V^{(l)}-c_A^{(l)}\gamma_5)\right\}\psi^{(l)}=0
\;\;, \end{equation}
where $l=e,\nu$ denotes the electron and neutrino case, respectively, 
and $c_{V,A}^{(e)}\equiv c_{V,A}$, $c_{V,A}^{(\nu )}\equiv 1$; the neutrino current has to be inserted into the 
electron equation and vice versa, as indicated by $J^{(l')}$ here. 
The V-A four-currents are defined by: 
\begin{equation}\label{current}
J_\mu^{(l)}\equiv\frac{G_F}{\sqrt 2}\langle :\bar\psi^{(l)}\gamma_\mu
(c_V^{(l)}-c_A^{(l)}\gamma_5)\psi^{(l)}:\rangle
\;\;, \end{equation} 
where the expectation value of the normal-ordered product refers to the 
ensemble characterizing the state of the system, which 
will be specified in more detail later. 
  
Introducing the Wigner functions, i.e. (4x4)-matrices with respect to the 
spinor indices which depend on space-time and four-momentum 
coordinates \cite{VGE87}: 
\begin{equation}\label{Wigner} 
W_{\alpha\beta}^{(l)}(x,p)\equiv\int\frac{\mbox{d}^4y}{(2\pi\hbar )^4}
\mbox{e}^{-ip\cdot y/\hbar}
\langle :\bar\psi_\beta^{(l)}(x+y/2)\psi_\alpha^{(l)}(x-y/2):\rangle
\;\;, \end{equation}
and with the $\hbar$-dependence made explicit, the currents of 
Eq.\,(\ref{current}) can be expressed as: 
\begin{equation}\label{currentWigner}
J_\mu^{(l)}(x)=\frac{G_F}{\sqrt 2}\;\mbox{tr}\int\mbox{d}^4p\;
\gamma_\mu (c_V^{(l)}-c_A^{(l)}\gamma_5)W^{(l)}(x,p) 
\;\;, \end{equation} 
with the trace refering to the spinor indices.      
 
Multiplying the Dirac equations (\ref{Dirac}) with the respective 
adjoint spinor and making use of 
Eqs.\,(\ref{current})-(\ref{currentWigner}), they can be converted 
to the Wigner representation. This yields the coupled electron and 
neutrino quantum transport equations:
\begin{equation}\label{transport}
\left (\gamma\cdot K-m^{(l)}\right )W^{(l)}(x,p)=
\exp (-\frac{i\hbar}{2}\partial_x\cdot\partial_p)  
J^{(l')}(x)\cdot\gamma (c_V^{(l)}-c_A^{(l)}\gamma_5)W^{(l)}(x,p)
\;\;, \end{equation} 
where $K_\mu\equiv p_\mu +\frac{i\hbar}{2}\partial_{x^\mu}$ and the 
partial derivative with respect to $x$ on the right-hand side acts 
only on the current $J^{(l')}$. If it were not for the V-A factor on the 
right-hand side, the structure of these equations would be analogous 
to Eq.\,(8) of Ref.\,\cite{EGVetal87} and could be analyzed accordingly.  
 
In order to proceed here, we employ 
the decomposition of the spinor Wigner functions \cite{VGE87}:
\begin{equation}\label{spindecomp}
W^{(l)}={\cal F}^{(l)}+i\gamma^5{\cal P}^{(l)}+\gamma^\mu {\cal V}^{(l)}_\mu 
+\gamma^\mu\gamma^5{\cal A}^{(l)}_\mu 
+\frac{1}{2}\sigma^{\mu\nu}{\cal S}^{(l)}_{\mu\nu}
\;\;, \end{equation} 
i.e., in terms of scalar, pseudoscalar, vector, axial vector, and 
antisymmetric tensor components: 
\begin{eqnarray}\label{scalar} 
{\cal F}^{(l)}(x,p)&\equiv&\frac{1}{4}\mbox{tr}\;W^{(l)}(x,p)
\;\;, \\ [1ex]
\label{pseudoscalar}  
{\cal P}^{(l)}(x,p)&\equiv&-\frac{1}{4}i\;\mbox{tr}\;\gamma^5 W^{(l)}(x,p)
\;\;, \\ [1ex]
\label{vector}
{\cal V}^{(l)}_\mu (x,p)&\equiv&\frac{1}{4}\mbox{tr}\;\gamma_\mu W^{(l)}(x,p)
\;\;, \\ [1ex]
\label{axialv} 
{\cal A}^{(l)}_\mu (x,p)&\equiv&\frac{1}{4}\mbox{tr}\;\gamma_5\gamma_\mu W^{(l)}(x,p)
\;\;, \\ [1ex]
\label{antisymt} 
{\cal S}^{(l)}_{\mu\nu}(x,p)&\equiv&\frac{1}{4}
\mbox{tr}\;\sigma_{\mu\nu}W^{(l)}(x,p)
\;\;, \end{eqnarray} 
which are real functions. Thus, for example, we obtain: 
\begin{equation}\label{currentdecomp}
J_\mu^{(l)}(x)=4\frac{G_F}{\sqrt 2}\int\mbox{d}^4p\;
[c_V^{(l)}{\cal V}^{(l)}_\mu (x,p)+c_A^{(l)}{\cal A}^{(l)}_\mu (x,p)] 
\;\;, \end{equation} 
using Eqs.\,(\ref{currentWigner}) and (\ref{vector}),\,(\ref{axialv}). 
Only these (axial) vector currents 
couple the transport equations (\ref{transport}).  

We introduce an abbreviation for the shift operator 
appearing in Eqs.\,(\ref{transport}), 
\begin{equation}\label{shiftcurrent}
{\cal J}^{(l)}_\mu\equiv 
\exp (-\frac{i\hbar}{2}\partial_x\cdot\partial_p)  
J^{(l)}_\mu (x)
=[ 
\cos (\frac{\hbar}{2}\partial_x\cdot\partial_p)
-i\sin (\frac{\hbar}{2}\partial_x\cdot\partial_p)]
J^{(l)}_\mu (x)
\equiv {\cal R}^{(l)}_\mu -i{\cal I}^{(l)}_\mu
\;\;, \end{equation}
where $\partial_x$ acts only on the current $J^{(l)}_\mu$, as 
before. Then, making use of the commutation  
and trace relations of the $\gamma$-matrices, we decompose 
Eqs.\,(\ref{transport}) in terms of the Wigner 
function components, Eqs.\,(\ref{scalar})-(\ref{antisymt}). 
Thus we obtain the set of coupled equations:   
\begin{eqnarray}\label{A}
K\cdot {\cal V}^{(l)}-m^{(l)}{\cal F}^{(l)} &=&
{\cal J}^{(l')}\cdot (c^{(l)}_V{\cal V}^{(l)}
+c^{(l)}_A{\cal A}^{(l)})
\;\;, \\ [2ex]
\label{B} 
iK\cdot {\cal A}^{(l)}+m^{(l)}{\cal P}^{(l)} &=& 
i{\cal J}^{(l')}\cdot (c^{(l)}_V{\cal A}^{(l)}
+c^{(l)}_A{\cal V}^{(l)})
\;\;, \end{eqnarray} 
\begin{eqnarray}
&\;&K_\mu {\cal F}^{(l)}-iK^\nu {\cal S}^{(l)}_{\mu\nu}-m^{(l)}{\cal V}^{(l)}_\mu 
\nonumber \\ &=& \label{C}
{\cal J}^{(l')\lambda}\left (g_{\mu\lambda}(c^{(l)}_V{\cal F}^{(l)}-ic^{(l)}_A{\cal P}^{(l)})
-ic^{(l)}_V{\cal S}^{(l)}_{\mu\lambda}
-\frac{1}{2}c^{(l)}_A\epsilon_{\mu\lambda\nu\nu '}{\cal S}^{(l)\nu\nu '}\right )
\;,\;\;\; \\ [2ex]
&\;&iK_\mu {\cal P}^{(l)}+\frac{1}{2}\epsilon_{\mu\lambda\nu\nu'}K^\lambda 
{\cal S}^{(l)\nu\nu'}-m^{(l)}{\cal A}^{(l)}_\mu 
\nonumber \\ &=& \label{D}
{\cal J}^{(l')\lambda}\left (g_{\mu\lambda}(ic^{(l)}_V{\cal P}^{(l)}-c^{(l)}_A{\cal F}^{(l)})
+\frac{1}{2}c^{(l)}_V\epsilon_{\mu\lambda\nu\nu '}{\cal S}^{(l)\nu\nu '}
+ic^{(l)}_A{\cal S}^{(l)}_{\mu\lambda}\right )
\;,\;\;\; \end{eqnarray}
\begin{eqnarray}
i(K_\mu {\cal V}^{(l)}_\nu -K_\nu {\cal V}^{(l)}_\mu ) 
-\epsilon_{\mu\nu\lambda\nu'}K^\lambda {\cal A}^{l\nu'} 
+m^{(l)}{\cal S}^{(l)}_{\mu\nu} &=&
{\cal J}^{(l')\lambda}\left (i[g_{\mu\lambda}
(c^{(l)}_V{\cal V}^{(l)}+c^{(l)}_A{\cal A}^{(l)})_\nu -\dots_{\;\mu\leftrightarrow\nu\;} ]\right . 
\nonumber \\ 
\label{E}
&\;&\;\;\;\;\;\;\;\;\;\;\left . 
-\epsilon_{\mu\nu\lambda\nu '}(c^{(l)}_V{\cal A}^{(l)}+c^{(l)}_A{\cal V}^{(l)})^{\nu '}
\right )
\;\;, \end{eqnarray}  
with $K_\mu$ as defined after Eq.\,(\ref{transport}).    

As is well known from other cases \cite{VGE87,EGV86,EGVetal87}, the real 
and imaginary parts of these coupled equations can be separated and eventually will thus lead 
to the proper transport equations in phase space and the generalizations  
of the mass-shell constraint, cf. Section 2.2\,. 
 
Furthermore, we observe that the left-hand sides of 
Eqs.\,(\ref{A})-(\ref{E}) formally coincide with Eqs.\,(5.7)-(5.11) of 
Ref.\,\cite{VGE87}. There, however, the corresponding operator $K^\mu$ 
for electrically charged particles necessarily incorporates the effects of the Lorentz force in the external field (Hartree) approximation. 

Due to the linearity of the 
Dirac equation with respect to the weak and electromagnetic interaction terms, i.e. with the derivative in Eq.\,(\ref{Dirac}) replaced according to the minimal 
coupling rule, 
$\partial_x^{\;\mu}\rightarrow\partial_x^{\;\mu}+ieA^\mu (x)$, it is 
straighforward to incorporate the electromagnetic interaction into 
Eqs.\,(\ref{A})-(\ref{E}) for the electron-positron case. 
Making use of the earlier QED results, this is achieved by the substitution:    
\begin{eqnarray}\label{K}
K^\mu&\equiv&\Pi^\mu +\frac{i\hbar}{2}\nabla^\mu 
\;\;, \\ [2ex] 
\label{Nabla}
\nabla^\mu&\equiv&\partial_x^{\;\mu} 
-ej_0(\textstyle{\frac{\hbar}{2}}
\partial_x\cdot\partial_p)
F^{\mu\nu}\partial_{p^\nu} 
\;\;, \\ [1ex]
\label{Pi}
\Pi^\mu&\equiv&p^\mu 
-e\textstyle{\frac{\hbar}{2}}j_1(\textstyle{\frac{\hbar}{2}}
\partial_x\cdot\partial_p)
F^{\mu\nu}\partial_{p^\nu} 
\;\;, \end{eqnarray} 
where $j_0$ and $j_1$ are the conventional spherical Bessel functions,  
cf. Eqs.\,(4.19)-(4.21) of Ref.\,\cite{VGE87};  
the derivatives $\partial_x$ in their arguments act only on the 
electromagnetic field strength tensor entering here, $F^{\mu\nu}(x)
\equiv\partial_x^{\;\mu}A^\nu (x)-\partial_x^{\;\nu}A^\mu (x)$. 
Our convention is that $e$ denotes the electron charge.

With the electromagnetic fields incorporated, we also need to include the 
Maxwell equation, 
\begin{equation}\label{Maxwell}
\partial_\mu F^{\mu\nu}(x)=J_{\mbox{em}}^\nu (x)
\equiv e\;\mbox{tr}\int\mbox{d}^4p\;\gamma^\nu W^{(e)}(x,p) 
\;\;, \end{equation} 
which consistently determines $F^{\mu\nu}$ in terms of the electromagnetic 
four-current $J_{\mbox{em}}^\nu$. However, an important remark is in order here.  
Together with Eqs.\,(\ref{K})-(\ref{Pi}) also the definition of 
the Wigner function (\ref{Wigner}) has to be modified. In order to preserve  
the gauge covariance of the equations, one has to include an  
appropriate electromagnetic phase factor (`Schwinger string')   \cite{VGE87,EGV86}. Since 
it will not appear explicitly in any of our further derivations 
or applications, it may presently suffice to keep this in mind.      

This completes the derivation of the coupled transport equations 
for a system of electrons, neutrinos, and electromagnetic fields 
in accordance with the Standard Model and in the collisionless (Vlasov) limit. 
 
\subsection{The Semiclassical Limit}
Our aim in this section is to extract the relevant semiclassical 
equations from the quantum transport equations which we obtained in the 
previous section, Eqs.\,(\ref{A})-(\ref{E}) in particular.   
Taking the explicit $\hbar$-dependence into account, which enters through the definitions of the shift and kinetic operators 
in Eqs.\,(\ref{shiftcurrent}) and (\ref{K})-(\ref{Pi}) 
respectively, it becomes 
obvious how to expand the equations in powers of $\hbar$. 
Since the leading terms of the real and imaginary parts of the equations start out with different powers, it is useful to separate them, 
similarly to what was previously done \cite{VGE87,EGV86,EGVetal87}. 
 
Furthermore, we presently simplify the set of equations by assuming 
a {\it spin saturated electron-positron plasma}, i.e. without the 
spin polarization effects which may be induced by strong magnetic fields, 
for example. Thus, for the $e^+e^-$ plasma, we have no pseudoscalar or axial vector 
densities, cf. Eqs.\,(\ref{spindecomp})-(\ref{antisymt}). 

Also, the Standard Model neutrino-antineutrino system
consists strictly only of left-handed neutrinos $\nu_L$ and a right-handed antineutrinos $\bar\nu_R$,  
if we appropriately neglect here their tiny (possibly finite) masses.
In this case, as we show in the Appendix, only the equal vector and axial vector densities 
contribute to the neutrino Wigner function,  
while all other densities vanish in the massless limit. 
 
These approximations serve as a working hypothesis for our 
study of the collective modes and their (in)stability in a 
supernova environment in Section 3\,. Eventually, however, the analysis  
of the complete coupled set of equations (\ref{A})-(\ref{E}) and (\ref{Maxwell}) should be performed, 
considering the presence or generation of strong magnetic fields during supernova explosions 
or other astrophysical processes \cite{Shukla98,Mikheev99,Wurtele99} (and references therein).  

\subsubsection{The Semiclassical $e^+e^-$ Transport Equations} 
Implementing ${\cal P}^{(e)}={\cal A}^{(e)}_\mu\equiv 0$ (spin saturation) and 
separating the real and imaginary parts of Eqs.\,(\ref{A})-(\ref{E}) 
with the help of Eqs.\,(\ref{shiftcurrent}) and (\ref{K}), 
we obtain for the $e^+e^-$ plasma the set of equations: 
\begin{eqnarray}
\label{Ar}
\Pi\cdot {\cal V}^{(e)}-m^{(e)}{\cal F}^{(e)}&=&0
\;\;, \\ [2ex]
\label{Ai}
\hbar\nabla\cdot {\cal V}^{(e)}&=&0
\;\;, \\ [2ex]
\label{Br} 
0&=&{\cal R}^{(\nu )}\cdot{\cal V}^{(e)}
\;\;, \\ [2ex]
\label{Bi} 
0&=&{\cal I}^{(\nu )}\cdot{\cal V}^{(e)}
\;\;, \end{eqnarray} 
\begin{eqnarray}
\label{Cr}
\Pi_\mu {\cal F}^{(e)}
+\frac{\hbar}{2}\nabla^\nu {\cal S}^{(e)}_{\mu\nu}-m^{(e)}{\cal V}^{(e)}_\mu &=& 
{\cal R}^{(\nu )\lambda}(g_{\mu\lambda}c_V{\cal F}^{(e)}
-\frac{1}{2}c_A\epsilon_{\mu\lambda\nu\nu '}{\cal S}^{(e)\nu\nu '})
-c_V{\cal I}^{(\nu )\lambda}{\cal S}^{(e)}_{\mu\lambda}
\;\;, \\ [2ex]
\label{Ci}
\frac{\hbar}{2}\nabla_\mu {\cal F}^{(e)}
-\Pi^\nu {\cal S}^{(e)}_{\mu\nu} &=& 
-{\cal I}^{(\nu )\lambda}(g_{\mu\lambda}c_V{\cal F}^{(e)}
-\frac{1}{2}c_A\epsilon_{\mu\lambda\nu\nu '}{\cal S}^{(e)\nu\nu '})
-c_V{\cal R}^{(\nu )\lambda}{\cal S}^{(e)}_{\mu\lambda}
\;\;, \\ [2ex]
\label{Dr}
\frac{1}{2}\epsilon_{\mu\lambda\nu\nu'}\Pi^\lambda 
{\cal S}^{(e)\nu\nu'} &=&
-{\cal R}^{(\nu )\lambda}(g_{\mu\lambda}c_A{\cal F}^{(e)}
-\frac{1}{2}c_V\epsilon_{\mu\lambda\nu\nu '}{\cal S}^{(e)\nu\nu '})
+c_A{\cal I}^{(\nu )\lambda}{\cal S}^{(e)}_{\mu\lambda}
\;\;, \\ [2ex]
\label{Di}
\frac{\hbar}{4}\epsilon_{\mu\lambda\nu\nu'}\nabla^\lambda 
{\cal S}^{(e)\nu\nu'} &=&
{\cal I}^{(\nu )\lambda}(g_{\mu\lambda}c_A{\cal F}^{(e)}
-\frac{1}{2}c_V\epsilon_{\mu\lambda\nu\nu '}{\cal S}^{(e)\nu\nu '})
+c_A{\cal R}^{(\nu )\lambda}{\cal S}^{(e)}_{\mu\lambda}
\;\;, \end{eqnarray}
\begin{eqnarray}
\label{Er}
-\frac{\hbar}{2}(\nabla_\mu {\cal V}^{(e)}_\nu -\nabla_\nu {\cal V}^{(e)}_\mu ) 
+m^{(e)}{\cal S}^{(e)}_{\mu\nu} &=&
c_V({\cal I}^{(\nu )}_\mu{\cal V}^{(e)}_\nu 
-{\cal I}^{(\nu )}_\nu{\cal V}^{(e)}_\mu )
-c_A\epsilon_{\mu\nu\lambda\nu '}{\cal R}^{(\nu )\lambda}{\cal V}^{(e)\nu '}
\;\;, \\ [2ex]
\label{Ei}
\Pi_\mu {\cal V}^{(e)}_\nu -\Pi_\nu {\cal V}^{(e)}_\mu  
 &=&
c_V({\cal R}^{(\nu )}_\mu{\cal V}^{(e)}_\nu 
-{\cal R}^{(\nu )}_\nu{\cal V}^{(e)}_\mu )
+c_A\epsilon_{\mu\nu\lambda\nu '}{\cal I}^{(\nu )\lambda}{\cal V}^{(e)\nu '}
\;\;, \end{eqnarray}  
where the constraints (\ref{Br}) and (\ref{Bi}), which resulted from 
Eq.\,(\ref{B}), where taken into account in Eqs.\,(\ref{Ar}) and 
(\ref{Ai}), which resulted from Eq.\,(\ref{A}); we used $c^{(e)}_{V,A}=c_{V,A}$\,, 
cf. Eqs.\,(\ref{Lint}),\,(\ref{Dirac}). 

We proceed to evaluate the limit $\hbar\rightarrow 0$ of 
the above system of equations. To begin with, we obtain  
from Eqs.\,(\ref{shiftcurrent}) that 
${\cal R}^{(l)}_\mu =J^{(l)}_\mu +O(\hbar^2)$ and 
${\cal I}^{(l)}_\mu =\frac{\hbar}{2}\partial_x\cdot\partial_pJ^{(l)}_\mu +O(\hbar^3)$ 
and from Eqs.\,(\ref{Nabla}),\,(\ref{Pi}) 
that $\Pi^\mu =p^\mu +O(\hbar^2)$ and 
$\nabla^\mu =\partial_x^{\;\mu}-eF^{\mu\nu}\partial_{p^\nu}+O(\hbar^2)$.  

Then, first of all, the vector density can formally be calculated 
from Eq.\,(\ref{Cr}): 
\begin{eqnarray}
{\cal V}^{(e)}_\mu &=&\frac{1}{m^{(e)}}\left (
(p_\mu -c_VJ^{(\nu )}_\mu ){\cal F}^{(e)}
+\frac{1}{2}c_A\epsilon_{\mu\lambda\nu\nu '}
J^{(\nu )\lambda}{\cal S}^{l\nu\nu '}\right .
\nonumber \\[1ex]
\label{CrSCl}
&\;&\;\;\;\;\;\;\;\left .
+\frac{\hbar}{2}(\partial_x^{\;\nu}-eF^{\nu\lambda}\partial_{p^\lambda})
{\cal S}^{(e)}_{\mu\nu}
+c_V\frac{\hbar}{2}\partial_x\cdot\partial_pJ^{(\nu )\lambda}
{\cal S}^{(e)}_{\mu\lambda}
\right )
+O(\hbar^2)
\;\;, \end{eqnarray}
where the right-hand side is to be evaluated consistently to first order;  
we recall that $\partial_x$ acts only on $J^{(\nu )}$ in the last term.
Similarly, we obtain from Eq.\,(\ref{Er}): 
\begin{eqnarray}
{\cal S}^{(e)}_{\mu\nu} &=&\left .\frac{1}{m^{(e)}}\right (
-c_A\epsilon_{\mu\nu\lambda\nu '}J^{(\nu )\lambda}{\cal V}^{(e)\nu '}
\nonumber \\[1ex]
\label{ErSCl}
&\;&\left .
+\frac{\hbar}{2}
[(\partial_{x^\mu}-eF_{\mu\lambda}\partial_p^{\;\lambda}){\cal V}^{(e)}_\nu 
-\dots_{\;\mu\leftrightarrow\nu\;}] 
+c_V\frac{\hbar}{2}\partial_x\cdot\partial_p
(J^{(\nu )}_\mu{\cal V}^{(e)}_\nu -J^{(\nu )}_\nu{\cal V}^{(e)}_\mu )\right )
+O(\hbar^2)
\;\;, \end{eqnarray}
with a contribution at $O(\hbar^0)$ in the absence of a 
pseudo-vector (or -scalar) density,
in distinction to the QED case of Ref.\,\cite{VGE87}.
Taking the limit $\hbar\rightarrow 0$, we solve   
Eqs.\,(\ref{CrSCl})-(\ref{ErSCl}) in terms of the 
scalar density ${\cal F}^{(e)}$ or, rather, the modified scalar density, 
\begin{equation}\label{modF}
\tilde f^{(e)}(x,p)\equiv\frac{{\cal F}^{(e)}(x,p)}
{1+(c_A/m^{(e)})^2J^{(\nu )}(x)\cdot J^{(\nu )}(x)}
\;\;. \end{equation}
The results are: 
\begin{eqnarray}\label{CrCl}
{\cal V}^{(e)}_\mu &=&\frac{1}{m^{(e)}}(p_\mu -c_VJ^{(\nu )}_\mu )
\tilde f^{(e)}
\;\;, \\ [1ex]
\label{ErCl}
{\cal S}^{(e)}_{\mu\nu} &=&\frac{c_A}{m^{(e)\;2}}
\epsilon_{\mu\nu\nu '\lambda}(p^{\nu '}-c_VJ^{(\nu )\nu '})J^{(\nu )\lambda}
\tilde f^{(e)}
\;\;, \end{eqnarray}
where we made use of the constraint (\ref{Br}), i.e. 
$J^{(\nu )}\cdot{\cal V}=0$ for $\hbar\rightarrow 0$, 
and conveniently added a term on the right-hand side of Eq.\,(\ref{ErCl})  
which vanishes identically. 
Thus, we find that in the semiclassical limit 
the spinor Wigner functions for the spin saturated system 
are completely determined by the scalar density, cf. Eq.\,(\ref{spindecomp}).
 
Next, using Eqs.\,(\ref{modF}) and (\ref{CrCl}), the Eq.\,(\ref{Ai})    
yields a transport equation for the scalar density: 
\begin{equation}\label{scaltransp}
(\partial_x^{\;\mu}-eF^{\mu\nu}\partial_{p^\nu})(p_\mu-c_VJ^{(\nu )}_\mu )
\tilde f^{(e)} 
=\left ((p_\mu -c_VJ^{(\nu )}_\mu )
(\partial_x^{\;\mu}-eF^{\mu\nu}\partial_{p^\nu})
-c_V(\partial_x\cdot J^{(\nu )})\right )\tilde f^{(e)}=0
\;\;, \end{equation} 
i.e. in the limit $\hbar\rightarrow 0$\,. Similarly, 
we obtain from Eq.\,(\ref{Ar}) 
together with Eq.\,(\ref{CrCl}) a  
constraint equation: 
\begin{equation}\label{massconstr}
\left ((p-c_VJ^{(\nu )})^2-m^{(e)\;2}(1+(c_A/m^{(e)})^2J^{(\nu )}
\cdot J^{(\nu )})\right )\tilde f^{(e)}=0
\;\;, \end{equation}
where we also used the constraint (\ref{Br}) in the 
form: 
\begin{equation}\label{BrCl}
(p-c_VJ^{(\nu )})\cdot J^{(\nu )}\tilde f^{(e)}=0
\;\;, \end{equation} 
which is appropriate in this limit. 
  
Clearly, the Eq.\,(\ref{massconstr}) 
demonstrates that it is the {\it kinetic momentum}, 
\begin{equation}\label{kinmom}
k^{(e)}_\mu\equiv p_\mu -c_VJ^{(\nu )}_\mu (x)
\;\;, \end{equation} 
which should be related to a classical mass-shell constraint. 
Therefore, we redefine the scalar density as a function of the 
kinetic momentum $k$, 
\begin{equation}\label{modFCl}
\tilde f^{(e)}(x,p)=\tilde f^{(e)}(x,k^{(e)}+c_VJ^{(\nu )})
\equiv f^{(e)}(x,k)
\;\;, \end{equation}
instead of the canonical momentum $p$; we will omit the superscript  
from $k^{(e)}$, since it is identical to the one of $f^{(e)}$ in the respective 
equations. This implies:
\begin{equation}\label{ptok}
\partial_x^{\;\mu}\tilde f^{(e)}|_p=\partial_x^{\;\mu}f^{(e)}|_k
-c_V(\partial_x^{\;\mu}J^{(\nu )}_\nu )\partial_k^{\;\nu}f^{(e)}
\;\;. \end{equation} 
   
For the redefined variable and scalar density function, 
the $e^+e^-$ {\it mass-shell constraint} follows: 
\begin{equation}\label{mass-shell}
\left (k^2-m^{(e)\;2}-c_A^{\;2}J^{(\nu )}
\cdot J^{(\nu )})\right )f^{(e)}=0
\;\;, \end{equation}
instead of Eq.\,(\ref{massconstr}). Furthermore, we finally obtain from 
Eq.\,(\ref{scaltransp}) the Vlasov type {\it transport equation} for 
the scalar $e^+e^-$ density: 
\begin{eqnarray}\label{Vlasov}
\left (k\cdot\partial_x
-c_Vk_\mu (\partial_x^{\;\mu}J^{(\nu )\nu})\partial_{k^\nu}
-c_V(\partial_x\cdot J^{(\nu )})
-ek_\mu F^{\mu\nu}\partial_{k^\nu}
\right )f^{(e)}&\;&
\nonumber \\ [1ex]
=\;\left (k\cdot\partial_x
-k_\mu (c_V[\partial_x^{\;\mu}J^{(\nu )\nu}-\partial_x^{\;\nu}J^{(\nu )\mu}]
+eF^{\mu\nu})\partial_{k^\nu})
\right )f^{(e)}&=&0
\;\;, \end{eqnarray} 
rewriting and using here the appropriate leading order in $\hbar$ form of the constraint (\ref{Bi}): 
\begin{equation}\label{BiCl}
(\partial_x^{\;\nu}J^{(\nu )\mu})\partial_{p^\nu}
(p_\mu -c_VJ^{(\nu )}_\mu )\tilde f^{(e)} 
=\left (k_\mu (\partial_x^{\;\nu}J^{(\nu )\mu})\partial_{k^\nu}
+(\partial_x\cdot J^{(\nu )})\right )f^{(e)}=0
\;\;. \end{equation}
In particular, we also employed Eq.\,(\ref{CrCl}), rewritten now simply as: 
\begin{equation}\label{vectordens}
{\cal V}^{(e)}_\mu =\frac{k_\mu}{m^{(e)}}f^{(e)}
\;\;. \end{equation}
We observe that the weak current-current interaction leads to an antisymmetric tensor coupling in the transport 
equation (\ref{Vlasov}), which is analogous to the electromagnetic 
field strength coupling. 
  
Furthermore, we remark that there are remaining equations of the set (\ref{Ar})-(\ref{Ei}) which we did not 
consider here, since the dynamics can be represented completely in terms of the scalar density ${\cal F}^{(e)}$, 
recall Eqs.\,(\ref{modF})-(\ref{ErCl}). Similarly as in the QED 
case of Ref.\,\cite{VGE87}, they could be shown to be satisfied identically to leading order 
in the $\hbar$-expansion, which we do not pursue here.
    
\subsubsection{The Semiclassical $\nu_L\bar\nu_R$ Transport Equations and Currents} 
For approximately massless neutrinos, with ${\cal V}^{(\nu )}_\mu ={\cal A}^{(\nu )}_\mu$ 
and ${\cal F}={\cal P}={\cal S}_{\mu\nu}\equiv 0$ (see Appendix), we obtain a much simpler set of equations from 
Eqs.\,(\ref{A})-(\ref{E}):
\begin{eqnarray}\label{neutrino1}
(K-2{\cal J}^{(e)})\cdot {\cal V}^{(\nu )}&=&0
\;\;, \\ [2ex]
\label{neutrino2}
i[(K-2{\cal J}^{(e)})_\mu {\cal V}^{(\nu )}_\nu -\dots_{\;\mu\leftrightarrow\nu\;} ]
-\epsilon_{\mu\nu\lambda\nu'}(K-2{\cal J}^{(e)})^\lambda {\cal V}^{(\nu )\nu'} 
&=&0
\;\;, \end{eqnarray}  
using $c_V^{(\nu )}+c_A^{(\nu )}=2$\,; here $K_\mu\equiv p_\mu +\frac{i\hbar}{2}\partial_{x^\mu}$\,. 
Separating real and 
imaginary parts, we expand the resulting equations in powers of $\hbar$\,: 
\begin{eqnarray}\label{neutrino3}
(p-2J^{(e)})\cdot {\cal V}^{(\nu )}
+O(\hbar^2)&=&0
\;\;, \\ [2ex]
\label{neutrino4}
(\partial_x+2\partial_x\cdot\partial_pJ^{(e)})\cdot {\cal V}^{(\nu )}
+O(\hbar^2)&=&0
\;\;, \\ [2ex]
\label{neutrino5}
[(p-2J^{(e)})_\mu {\cal V}^{(\nu )}_\nu -\dots_{\;\mu\leftrightarrow\nu\;} ]
+O(\hbar )&=&0
\;\;, \\ [2ex]
\label{neutrino6}
\epsilon_{\mu\nu\lambda\nu'}(p-2J^{(e)})^\lambda {\cal V}^{(\nu )\nu'} 
+O(\hbar )&=&0
\;\;, \end{eqnarray}  
thus proceeding similarly as in the case of the electron-positron plasma up to this point. 
  
However, now it is obvious that the following Ansatz immediately solves  
Eqs.\,(\ref{neutrino5}) and (\ref{neutrino6}): 
\begin{equation}\label{fnutilde}
{\cal V}^{(\nu )}_\mu\equiv (p-2J^{(e)})_\mu\tilde f^{(\nu )}
\;\;, \end{equation}
with a scalar function $\tilde f$ of the phase space variables $x,p$\,.
Furthermore, to leading order in $\hbar$\,, it converts Eq.\,(\ref{neutrino1}) into the mass-shell constraint:  
\begin{equation}\label{neutrino11} 
(p-2J^{(e)})^2\tilde f^{(\nu )}=0
\;\;, \end{equation} 
which demonstrates that it is the {\it kinetic momentum}, $k_\mu\equiv p_\mu -2J^{(e)}(x)$\,, 
which is to be on-shell here.  
 
Performing analogous steps as in Eqs.\,(\ref{kinmom})-(\ref{ptok}) before, redefining 
$\tilde f^{(\nu )}(x,p)\equiv f^{(\nu )}(x,k)$ in particular, we obtain directly the $\nu_L\bar\nu_R$ {\it mass-shell constraint}: 
\begin{equation}\label{numass-shell}
k^2f^{(\nu )}=0
\;\;, \end{equation}  
and from Eq.\,(\ref{neutrino4}) the Vlasov type {\it transport equation} for the $\nu_L\bar\nu_R$ density function: 
\begin{equation}\label{nuVlasov} 
\left (k\cdot\partial_x
-2k_\mu [\partial_x^{\;\mu}J^{(e)\nu}-\partial_x^{\;\nu}J^{(e)\mu}]\partial_{k^\nu}
\right )f^{(\nu )}=0
\;\;, \end{equation}
which may be compared to Eqs.\,(\ref{mass-shell}) and (\ref{Vlasov}) of the electron-positron plasma. 

In order to complete the set of coupled classical transport and 
constraint equations (\ref{mass-shell}), (\ref{Vlasov}), (\ref{numass-shell}), and (\ref{nuVlasov}), we have to 
reconsider the four-currents entering here and into the Maxwell equation 
(\ref{Maxwell}) in the limit $\hbar\rightarrow 0$\,. 

Implementing the spin saturation, in particular ${\cal A}^{(e)}_\mu\equiv 0$\,, and using  Eqs.\,(\ref{CrCl}) and (\ref{modFCl}), we obtain the {\it weak} $e^+e^-$ {\it current}: 
\begin{equation}\label{eweakcurrent}
J^{(e)}_\mu (x)=4\frac{G_F}{\sqrt 2}\frac{c_V}{m^{(e)}}
\int\mbox{d}^4p\; (p_\mu -c_VJ^{(\nu )}_\mu (x))\tilde f^{(e)}(x,p)
=4\frac{c_VG_F}{\sqrt 2}
\int\mbox{d}^4k\; \frac{k_\mu}{m^{(e)}}f^{(e)}(x,k)
\;\;, \end{equation}
cf. Eq.\,(\ref{currentdecomp}), with $c^{(e)}_V=c_V$\,. Similarly, 
the {\it electromagnetic current} assumes the form:
\begin{equation}\label{electrocurrent}
J^\mu_{\mbox{em}}(x)=4e\int\mbox{d}^4k\; \frac{k_\mu}{m_e}f^e(x,k)
\;\;, \end{equation}
cf. Eq.\,(\ref{Maxwell}). 
Finally, the {\it weak} $\nu_L\bar\nu_R$ {\it current} is:  
\begin{equation}\label{nuweakcurrent}
J^{(\nu )}_\mu (x)=8\frac{G_F}{\sqrt 2}
\int\mbox{d}^4p\; (p_\mu -2J^{(e)}_\mu (x))\tilde f^{(\nu )}(x,p)
=8\frac{G_F}{\sqrt 2}
\int\mbox{d}^4k\; k_\mu f^{(\nu )}(x,k)
\;\;, \end{equation}
using once more ${\cal V}^{(\nu )}_\mu ={\cal A}^{(\nu )}_\mu$\,, $c_V^{(\nu )}+c_A^{(\nu )}=2$\,, as well as  Eq.\,(\ref{fnutilde}). 
  
The closed set of four coupled mass-shell constraint and transport equations, with the 
currents determined by scalar (density) functions, along with the Maxwell 
equation present the final result 
of our derivation of the semiclassical nonequilibrium transport theory
of neutrinos and electrons. It incorporates their antiparticles as 
discussed in more detail for the QED case in Ref.\,\cite{VGE87}, as 
well as electromagnetic fields, assuming an $e^+e^-$-spin saturated system in   
the mean field dominated regime.   
 
We note that the structure of our {\it final} closed set of equations could  
essentially be anticipated from purely classical kinetic theory considerations, 
as previously observed \cite{Bento99,EGV86}. On the other hand, for the study of spin-polarization 
or strong magnetic field effects and higher order quantum corrections, we must go back to our previous set of 
Eqs.\,(\ref{A})-(\ref{E}) of Section 2.1\,.

\section{Linear Response Analysis and (Un)Stable Collective Modes}
Presently, we apply the transport theory of Section 2\,, in order 
to derive the semiclassical dispersion relations of collective modes of 
a neutrino-electron system in general (Section 3.1). 

In Section 3.2 we specialize our results for the Type\,II supernova scenario. The relevant distribution functions are introduced 
in Section 3.2.1 and the necessary response functions calculated 
in Section 3.2.2\,. 

In Section 3.2.3 we evaluate the dispersion relations for various 
collective modes. We determine their (in)stability 
properties in the neutrino `beam' plus electron 
`plasma sphere' system formed during a Type\,II supernova explosion. 
The final Section 3.2.4 is devoted to a discussion of the validity of the 
approximations used. 

\subsection{The Linear Response Theory for Neutrino-Electron Systems} 
The behavior of collective modes, in particular, the onset of 
instabilities, is determined by the evolution of small 
perturbations of a generic set of stationary distribution functions, 
which may be caused by scattering interactions, for example. Therefore, 
we write the scalar density distributions in the form:
\begin{equation}\label{pertdensity}
f^{(l)}(x,k)=f^{(l)}_S(k)+\delta f^{(l)}(x,k)
\;\;, \end{equation}
where $f^{(l)}_S$ denotes the assumed {\it homogeneous} four-momentum dependent
solutions of the mass-shell and transport equations and 
$\delta f^{(l)}$ an initially {\it small} perturbation. 
This assumption of homogeneity greatly simplifies the subsequent 
analysis and describes a sufficiently large `free-streaming'  
electron-neutrino system.   
 
The weak currents $J^{(l)}_\mu$ determined by $f^{(l)}_S$, cf.   
Eqs.\,(\ref{eweakcurrent}) and (\ref{nuweakcurrent}), are homogeneous and the antisymmetric 
tensors which enter the transport equations (\ref{Vlasov}) and (\ref{nuVlasov}),
\begin{equation}\label{Gmunu}
G^{(l)\mu\nu}\equiv c^{(l)}[\partial_x^{\;\mu}J^{(l')\nu}-\partial_x^{\;\nu}J^{(l')\mu}]
\;\;, \end{equation}
vanish in this case; from here on $c^{(e)}\equiv c_V$ and $c^{(\nu )}\equiv 2$\,. 
Furthermore, assuming an {\it isotropic on-shell} electron-positron distribution, 
\begin{equation}\label{e0density}
f^{(e)}_S(k)\equiv
\delta [k^2-m^{(e)\;2}-c_A^{\;2}J^{(\nu )}\cdot J^{(\nu )} ] 
f^{(e)}(k^0,|\vec k|)
\;\;, \end{equation}
cf. Eq.\,(\ref{mass-shell}), 
it follows that the corresponding electromagnetic four-current (\ref{electrocurrent}) vanishes, if we additionally assume a 
neutralizing  background charge or approximately 
equal densities of electrons and positrons, depending on the 
circumstances. 
Consistently we set $F^{\mu\nu}\equiv 0$\,, i.e.,  
considering a situation without external electromagnetic fields. 

Indeed, then, the initial on-shell distributions $f^{(\nu )}_S(k)$ and 
$f^{(e)}_S(k)$ are {\it stationary} in the absence of collisions.     
They will be further specified shortly.  
   
Linearizing the transport equations (\ref{Vlasov}) and (\ref{nuVlasov}) with respect to the 
small perturbations $\delta f^{(l)}$, we obtain for the electrons: 
\begin{equation}\label{deltaeVlasov}
ik\cdot q\;\delta f^{(e)}(q,k)+k_\mu\left(\delta G^{(e)\mu\nu}(q)+
e\delta F^{\mu\nu}(q)\right )\partial_{k^\nu}f^{(e)}_S(k)=0
\;\;, \end{equation}
where we introduced the Fourier transform for any function $g$ of the 
space-time coordinates, 
$g(x)\equiv(2\pi )^{-4}\int\mbox{d}^4q\exp (-iq\cdot x)g(q)$. Here
$\delta G^{(e)}$ and $\delta F$ denote the weak and electromagnetic 
tensors induced by the perturbations $\delta f^{(\nu )}$ and $\delta f^{(e)}$,  
respectively. From Eq.\,(\ref{Gmunu}) we obtain:
\begin{equation}\label{Gmunuq}
\delta G^{(l)\mu\nu}(q)=-ic^{(l)}(q^\mu g^{\nu\lambda} -q^\nu g^{\mu\lambda})
\delta J^{(l')}_\lambda
\;\;. \end{equation}
Furthermore, solving the Maxwell equation (\ref{Maxwell}) with the retarded 
boundary condition (damping in the infinite past), we obtain: 
\begin{equation}\label{Maxwellsol}
e\delta F^{\mu\nu}(q)=\frac{ie}{q^2+i\epsilon q^0}(q^\mu g^\nu_\lambda
-q^\nu g^\mu_\lambda )\delta J^\lambda_{\mbox{em}}
=i\frac{\sqrt 2e^2}{c_VG_F}
\frac{1}{q^2+i\epsilon q^0}(q^\mu g^{\nu\lambda}
-q^\nu g^{\mu\lambda})\delta J^{(e)}_\lambda
\;\;, \end{equation} 
where $\epsilon\rightarrow 0^+$, and where we used Eqs.\,(\ref{eweakcurrent}) and (\ref{electrocurrent}), in order to express the (conserved) electromagnetic current fluctuation in terms of its weak counterpart.

Implementing the retarded boundary condition, i.e. the `Landau prescription' \cite{LifshitzP81}, the electron transport equation (\ref{deltaeVlasov}) is solved by: 
\begin{equation}\label{deltaeVlasovSol}
\delta f^{(e)}(q,k)=
\frac{k_\mu\left(\delta G^{(e)\mu\nu}(q)+
e\delta F^{\mu\nu}(q)\right )\partial_{k^\nu}}
{-i(k\cdot q+i\epsilon k^0)}
f^{(e)}_S(k)
\;\;. \end{equation}
Similarly, the perturbation of the stationary neutrino distribution is 
determined by:
\begin{equation}\label{deltanVlasovSol}
\delta f^{(\nu )}(q,k)=
\frac{k_\mu\delta G^{(\nu )\mu\nu}(q)
\partial_{k^\nu}}
{-i(k\cdot q+i\epsilon k^0)}
f^{(\nu )}_S(k)
\;\;. \end{equation}
Obviously, Eqs.\,(\ref{deltaeVlasovSol}) and (\ref{deltanVlasovSol}) are  
coupled to each other via Eqs.\,(\ref{Gmunuq}).    

We proceed by introducing the response functions:
\begin{equation}\label{response}
M^{(l)\lambda\rho}(q)\equiv 4\int\mbox{d}^4k\;\frac{k^\lambda}{m^{(l=e)}}
\frac{1}{k\cdot q+i\epsilon k^0}
(k\cdot q\partial_k^{\;\rho}-k^\rho q\cdot\partial_k)f^{(l)}_S(k)
\;\;, \end{equation} 
which will be calculated for specific choices of the stationary 
distributions $f^{(l)}_S$ shortly; the factor $1/m^{(l=e)}$ is meant to apply only 
in the $e^+e^-$ case and to be replaced by 1 for the (approximately) massless $\nu_L\bar\nu_R$ case.    

Making use of the response functions, 
we multiply Eqs.\,(\ref{deltaeVlasovSol}) and (\ref{deltanVlasovSol}) 
by the appropriate factors, cf. Eqs.\,(\ref{eweakcurrent}) and (\ref{nuweakcurrent}),
and integrate over $\mbox{d}^4k$, in order to obtain a closed set of algebraic equations: 
\begin{eqnarray}\label{deltaecurrent}
\delta J^{(e)\lambda}(q)&=&M^{(e)\lambda\rho}(q)\left (
\frac{c_V^{\;2}G_F}{\sqrt 2}\delta J^{(\nu )}_\rho (q)
-\frac{e^2}{q^2+i\epsilon q^0}\delta J^{(e)}_\rho (q)\right )
\;\;, \\ [2ex] 
\label{deltancurrent}
\delta J^{(\nu )\lambda}(q)&=&\frac{4G_F}{\sqrt 2}
M^{(\nu )\lambda\rho}(q)
\delta J^{(e)}_\rho (q)
\;\;, \end{eqnarray}
where the tensor fluctuations $\delta G^{(l)}$ and $\delta F$ were eliminated 
with the help of Eqs.\,(\ref{Gmunuq}) and (\ref{Maxwellsol}), respectively.    
Inserting the second into the first equation, the final result is: 
\begin{equation}\label{deltaecurrentfinal}
{\cal M}^{(e)\lambda\rho}(q)\delta J^{(e)}_\rho (q)\equiv
\left [g^{\lambda\rho}+
M^{(e)\lambda\sigma}(q)\left (
\frac{e^2}{q^2+i\epsilon q^0}g^\rho_\sigma
-2c_V^{\;2}G_F^{\;2}
g_{\sigma\tau}M^{(\nu )\tau\rho}(q)
\right )\right ]\delta J^{(e)}_\rho (q)=0
\;\;. \end{equation}
The solvability condition of this vector equation determines 
the {\it dispersion relation} for the 
perturbations of the stationary electron-positron distribution: 
\begin{equation}\label{electrondispersion}
\mbox{Det}\;{\cal M}^{(e)}(q)=0
\;\;, \end{equation}
where ${\cal M}^{(e)}$ is a $4\times 4$-matrix in the 
Lorentz indices. Analogously one obtains the dispersion relation for  
the neutrino case, which we do not pursue. 

A final remark is in order here. From the structure of Eq.\,(\ref{deltaecurrentfinal}), particularly the generically 
small weak coupling  
term compared with the electromagnetic one, it is natural to 
expect that the neutrinos can only influence the resulting dispersion 
relations noticeably, if their response function shows rather 
singular behavior. Furthermore, in this case, weak and electromagnetic 
interactions presumably will mix in the corresponding collective modes, 
due to the products involved, for example, 
in Eqs.\,(\ref{deltaecurrentfinal}) and (\ref{electrondispersion}).
This will be studied in the following sections with the 
application to a supernova scenario. 
      
\subsection{The Supernova Two-Stream Scenario}  
The above results are fairly general and need to be specialized 
according to the physical nature of the stationary 
distributions as well as of their potentially unstable perturbations. 
We shall now study the idealized situation where an electrically neutral finite temperature electron-positron 
plasma is hit by a neutrino-antineutrino beam. (Anti-)neutrinos are radiated from 
the neutrino sphere, move approximately radially outwards, and 
interact {\it collectively} with the electron-positron plasma sphere forming  
the `radiation bubble' in Bethe's supernova scenario \cite{RECENT1}. As 
before, we derive our results in the collisionless limit.  

Typically,  
it is assumed that a short-lived, but intense neutrino flux 
($3\times 10^{29}\;\mbox{W/cm}^2$, total integrated luminosity up to  
the order of several $10^{53}$\,erg\,) with an approximately thermal spectrum corresponding to a temperature $T_\nu\approx 1\dots 10$\,MeV  
is released from the collapsing core and interacts at a distance of about $30\dots 300$\,km from the center with the surrounding 
moderately relativistic electron plasma of (charge) 
density $n_e\stackrel{<}{\sim} 10^{30}\,\mbox{cm}^{-3}$ and 
temperature $T_e\stackrel{>}{\sim} 0.5$\,MeV; 
here the uncertainties mostly reflect differing scenarios 
considered in this context \cite{Cooperstein88,RECENT1,RECENT2,Bingham94,Bento99}.  
We will study a corresponding set of parameters, following the discussion of the radiation bubble 
by Bethe \cite{RECENT1}. 
  
For the above (optimistic charge) density and temperature,   
the electron-positron plasma is nondegenerate, with an estimated chemical potential $\mu_e<\pi T$\,. 
This is indirectly supported by Bethe's 
results, see in particular Sections VI.\,E-G of Ref.\,\cite{RECENT1}, which demonstrate the dilute character 
of matter in the radiation bubble - the energy or entropy density of the `radiation' (i.e. of photons plus pairwise 
produced electrons and positrons) is more than a factor $10^2$ higher than that of nucleons in the bubble. 
Therefore, any background charge contamination by protons must be small here, and correspondingly the net 
charge of electrons over positrons neutralizing the plasma. Hence we may neglect  
the finite electron chemical potential in a first approximation.   

\subsubsection{Distribution Functions}     
The following stationary electron-positron distribution will now be 
considered, cf. Eq.\,(\ref{e0density}):
\begin{equation}\label{Te0density} 
f^{(e)}_S(k)=(2\pi )^{-3}m^{(e)}
\delta [k^2-m^{(e)\;2}-c_A^{\;2}J^{(\nu )}\cdot J^{(\nu )}]
\left( \Theta (k^0)F(k^0/T_e)+\Theta (-k^0)F(-k^0/T_e)\right )
\;\;, \end{equation}
where $F(x)\equiv (\mbox{e}^x+1)^{-1}$, and where $T_e$ 
denotes their temperature. When $J^{(\nu )}=0$\,,  Eq.\,(\ref{Te0density}) 
describes the $e^+e^-$ blackbody radiation (omitting the 
vacuum contribution). We remark that antiparticles are represented as fermions with negative 
four-momentum here \cite{VGE87}.

Concerning the emission from the neutrino sphere, we 
neglect its collective flow relative to the  
electron-positron plasma sphere, or vice versa. However, it is important to incorporate the dilution and angular squeezing effects due to the spherical geometry. Thus, we assume the following stationary 
(approximately massless) neutrino-antineutrino distribution:
\begin{equation}\label{Tn0density} 
f^{(\nu )}_S(k)=\frac{1}{2}(2\pi )^{-3} 
\delta [k^2]
\left( \Theta (k^0)\Theta (\theta_{max}-\theta_{\vec n,\vec k})
F([k^0-\mu_\nu ]/T_\nu )
+\dots\; k,\mu_\nu \rightarrow -k,-\mu_\nu 
\right )
\;, \end{equation}
where $\mu_\nu$ denotes their chemical potential and only 
{\it one} ($\nu_L$ or $\bar\nu_R$) spin state is 
taken into account. The additional $\Theta$-function, implementing the radial (`outward') 
unit vector $\vec n$, accounts for the 
finite opening angle $\theta_{\vec n,\vec k}$ between neutrino momenta and the radial direction. The maximal opening angle is determined by  $\sin\theta_{max}=R/r$, where $r$ denotes the distance from the center of the neutrino sphere of radius $R$. 
The usual dilution factor, $d\equiv (R/r)^2=\sin^2\theta_{max}$, does not appear explicitly, 
but is recovered in the calculation of, for example,  
the energy flux from the neutrino sphere based on $f_S^{(\nu )}$. 

We remark that 
the neutrino distribution is not necessarily uniform within 
the cone defined by $\theta_{max}$. 
It may vary considerably, depending on the emission characteristics of the neutrino sphere. Thus, the distribution of Eq.\,(\ref{Tn0density}) may  
represent an opening angle average; the (ir)relevance 
of the sharp $\Theta$-function cut-off will be discussed in the final 
subsection 3.2.4\,. Furthermore, it is {\it not} a global solution of the {\it spherical} free-streaming problem.  
However, our approximate spatially homogeneous distribution, with 
the parametric dependence on $R/r$, is sufficient for the  
study of collective modes with a characteristic wavelength very much less than $R$, even though we are interested in the long-wavelength limit with 
respect to the microscopic scales.

We omit the  
$\mu$- and $\tau$-neutrinos at present  
which have a considerably weaker effective coupling, see  
Eq.\,(\ref{deltaecurrentfinal}) together with the remarks 
after Eq.\,(\ref{Lint}); using $\sin^2\theta_W\approx 0.23$, we 
have $c_V\approx 0.96\; (-0.04)$ for the electron ($\mu,\tau$-) neutrinos. Furthermore, their chemical potential 
vanishes. For the electron 
(anti-)neutrinos, which carry about 4/10 of the 
total neutrino energy flux, we adopt Bethe's estimate which 
yields $\eta_\nu\equiv\mu_\nu /T_\nu =0.29$ \cite{RECENT1}.   

Next, we proceed to calculate the energy-momentum tensor, similarly 
as in Ref.\,\cite{VGE87}, for a stationary free 
neutrino-antineutrino distribution: 
\begin{equation}\label{nTmunu}
T_{\mu\nu}^{(\nu )}(x)=\mbox{tr}\int \mbox{d}^4k\;k_\nu\gamma_\mu W^{(\nu )}(x,k)
=4\int\mbox{d}^4k\;k_\nu {\cal V}^{(\nu )}_\mu (x,k)
\;\;, \end{equation}
i.e. in terms of the vector density, cf. Eq.\,(\ref{vector}). 
Employing Eq.\,(\ref{fnutilde}) and projecting on the `outward' 
momentum direction, we obtain the electron-$\nu_L\bar\nu_R$ energy flux 
corresponding to the homogeneous equilibrium distribution 
(\ref{Tn0density}): 
\begin{eqnarray}
T^{(\nu )0i}n^i&=&4\int\mbox{d}^4k\;k^0 k^z 
f_S^{(\nu )}(k)
= \frac{d}{8\pi^2}\int_0^\infty\mbox{d}k\; k^3 
\left( F([k-\mu_\nu ]/T_\nu )+F([k+\mu_\nu ]/T_\nu )\right )
\nonumber \\ [1ex]
\label{Eflux}
&=&\frac{7\pi^2}{480}dT_\nu^{\;4}
\left (1+\frac{30}{7\pi^2}\eta_\nu^{\;2}+\frac{15}{7\pi^4}\eta_\nu^{\;4}\right )
=\frac{4}{10}\cdot\frac{dL}{\pi R^2}
\;\;, \end{eqnarray}
choosing $\vec n=(0,0,1)$, and  
where $L$ denotes the {\it total} neutrino-plus-antineutrino luminosity. 
The integral is evaluated exactly with the help of a formula from 
Ref.\,\cite{EGR1980}. Correction terms involving powers of $m^{(\nu )}/(\eta_\nu T_\nu)$ would be 
completely negligible for temperatures in the MeV range and a typical neutrino mass (much) 
less than 1\,eV. The last equality in Eq.\,(\ref{Eflux}) 
provides the relation between temperature and radius 
of the neutrino sphere, given its luminosity $L$ \cite{RECENT1}.
 
Similarly, we obtain from Eq.\,(\ref{nuweakcurrent}) by direct 
calculation: 
\begin{equation}\label{Jnumu}
J^{(\nu )\mu}=8\frac{G_F}{\sqrt 2}\int\mbox{d}^4k\;
k^\mu f^{(\nu )}_S(k)= 
\frac{G_F}{\sqrt 2}
\frac{dT_\nu^{\;3}}{12}\left (
\eta_\nu +\frac{1}{\pi^2}\eta_\nu^{\;3}\right )\xi^\mu
\;\;, \end{equation} 
where $\xi^\mu\equiv (2/[1+\cos\theta_{max}],0,0,1)$\,.  
As expected, the neutrino current components are very small, 
since for temperatures of about 10\,MeV we have 
$G_FT_\nu^{\;3}\approx 10^{-8}$\,MeV. Therefore, the corresponding 
term $\propto J^{(\nu )}\cdot J^{(\nu )}$ in the expression for the stationary electron-positron distribution, cf. Eq.\,(\ref{Te0density}), 
can be safely neglected henceforth. 

\subsubsection{Response Functions}  
After specifying the unperturbed stationary 
electron and neutrino distributions, $f_S^{(e)}$ and $f_S^{(\nu )}$ respectively, 
we calculate the response functions $M^{(e)\lambda\rho}$ and $M^{(\nu )\lambda\rho}$ defined in Eq.\,(\ref{response}). For the following 
calculations it is convenient to perform a partial integration, which 
yields: 
\begin{equation}\label{response1} 
M^{(l)\lambda\rho}(q)=4\int\mbox{d}^4k\;\frac{f_S^{(l)}(k)}{m^{(l=e)}}
\left (-g^{\lambda\rho}+\frac{q^\lambda k^\rho +q^\rho k^\lambda}
{k\cdot q+i\epsilon k^0}-\frac{q^2k^\lambda k^\rho}
{(k\cdot q+i\epsilon k^0)^2}\right )
\;\;, \end{equation} 
which is now obviously symmetric and {\it transverse}, $q_\lambda M^{(l)\lambda\rho}(q)=0$. Thus, the current fluctuations are 
properly conserved, $q_\lambda\delta J^{l\lambda}(q)=0$, cf. 
Eqs.\,(\ref{deltaecurrent})-(\ref{deltancurrent}).  
 
Beginning with the electron case, the calculation is facilitated by 
recalling that the distribution function $f_S^{(e)}$, Eq.\,(\ref{Te0density}), 
is isotropic with respect to the three-momentum components. Therefore, 
the spatial part of the response function can be decomposed into a 
transverse and a longitudinal part, 
\begin{equation}\label{decomposition}
M^{(e)ij}(q)\equiv\left (\delta^{ij}-\frac{q^i q^j}{\vec q^2}
\right )M^{(e)}_T(q)+\frac{q^i q^j}{\vec q^2}M^{(e)}_L(q)
\;\;. \end{equation}
Defining the {\it electric} (Debye) {\it screening mass}, 
\begin{equation}\label{Dmass}
m_D^{2}\equiv \frac{4e^2}{\pi^2}\int_0^\infty\mbox{d}k\; k
F(k/T_e)=\frac{1}{3}e^2T_e^{\;2}
\;\;, \end{equation}
the results of a standard calculation for the {\it electron-positron  response function} are: 
\begin{eqnarray}\label{M00}
-e^2M^{(e)00}(q^0,\vec q)&=&m_D^{2}
\left [1-\frac{1}{2}\frac{\omega}{q}
\left (\ln\left |\frac{q+\omega }{q-\omega }\right |
-i\pi\Theta (q-\omega )\right )\right ] 
\;\;, \\ [1ex]
\label{M0i} 
M^{(e)0i}(q^0,\vec q)&=&M^{(e)i0}(q^0, \vec q)\;=\;
\frac{q^0q^i}{q^2}M^{(e)00}(q^0,\vec q)
\;\;, \\ [1ex]
\label{ML} 
M^{(e)}_L(q^0,\vec q)&=&\frac{\omega ^2}{q^2}M^{(e)00}(q^0,\vec q)
\;\;, \\ [1ex]
\label{MT} 
e^2M^{(e)}_T(q^0,\vec q)&=&
\frac{1}{2}m_D^{2}\frac{\omega^2}{q^2}
\left [1-\frac{1}{2}\left (\frac{\omega}{q}-\frac{q}{\omega}\right )
\left (\ln\left |\frac{q+\omega}{q-\omega}\right |
-i\pi\Theta (q-\omega )\right )\right ] 
\;\;, \end{eqnarray}
where we implemented the relativistic limit and neglected correction 
terms in powers of $m_e/T_e$; here we    
simplified the notation by introducing $\omega\equiv |q^0|$ and 
$q\equiv |\vec q|$. 
We note the appearance of the imaginary 
parts which, in general, are responsible for Landau damping \cite{LifshitzP81,Uli85}. 
These kinetic theory results completely 
agree with the perturbative one-loop evaluation of the QED polarization
tensor in the high-temperature limit \cite{Kapusta}, a correspondence which 
was has been observed in many other cases, e.g., \cite{EGV86}.   
  
In a fully realistic calculation, corrections due to the finite ratio of $m_e/T_e\stackrel{<}{\sim}1$ should also be 
considered. We neglect them in our present work, since the appropriate electron temperature is not precisely known 
in this context. Furthermore, unfortunately, this would necessitate numerical calculations 
where the transparency of the analytical results presented here would be lost. On the one hand, it seems 
unlikely that the additional mass scale can qualitatively change any of our conclusions, since it is well separated 
from all the plasma scales entering in the following. However, for particular effects, e.g. proper Landau damping 
\cite{LifshitzP81,Uli85}, a finite mass may be crucial (cf. footnote following the discussion after 
Eq.\,(112)). Thus, proper `neutrino Landau damping' has been discussed in more detail by Silva {\it et al.} recently \cite{Bingham94}.             

Next, we turn to the calculation of the neutrino response function 
$M^{(\nu )\lambda\rho}$. It is more involved due to the preferred direction of propagation, which enters here through the dependence of 
the stationary distribution (\ref{Tn0density}) on the `radial' unit vector 
$\vec n$. In order to facilitate our task, we consider two 
cases separately, depending on the orientation of the wave vector 
$\vec q$ with respect to $\vec n$: 
$\vec q\parallel\vec n$ ({\bf Case I}) and $\vec q\perp\vec n$ 
({\bf Case II}). 
We recall that $\vec q$ determines the direction of propagation of the  
collective excitations of the electron-positron plasma, especially in the presence of the neutrino flux. 
   
\underline{{\bf Case I}.} Here we expect a response function with a formal 
structure generalizing the familiar results of 
Eqs.\,(\ref{M00})-(\ref{MT}), since the geometry determining the 
essential angular integrations is identical to the 
previous case. Therefore, a tensor decomposition into transverse and 
longitudinal parts analogous to Eq.\,(\ref{decomposition}) still applies.   
However, the maximal opening 
angle $\theta_{max}$ between neutrino momenta and the `radial' 
direction limits the azimuthal angle $\theta$, e.g. in the 
vector decomposition, 
\begin{equation}\label{kdecomp}
\vec k=\vec qq^{-1}k\cos\theta +\vec k_\perp 
\;\;, \end{equation} 
with $k\equiv |\vec k|$, 
which is conveniently employed after converting the integral of Eq.\,(\ref{response1}) 
to the corresponding threedimensional (on-shell) form. 

Furthermore, 
instead of the Debye mass of Eq.\,(\ref{Dmass}), we  
introduce the {\it weak thermal mass}:  
\begin{equation}\label{Wmass}
m_w^{2}\equiv\frac{2c_V^{\;2}}{\pi^2}\int_0^\infty\mbox{d}k\;k
\left (F([k-\mu_\nu ]/T_\nu )+F([k+\mu_\nu ]/T_\nu )\right )
=\frac{1}{3}c_V^{\;2}T_\nu^{\;2}\left (1+\frac{3}{\pi^2}\eta_\nu^{\;2}
\right )
\;\;, \end{equation}
which takes the finite chemical potential of the neutrinos into 
account. Here we applied the ultrarelativistic limit discussed 
before, as well as the appropriate integral formula from 
Ref.\,\cite{EGR1980}.   

Then, we obtain the components of the {\it neutrino-antineutrino response function} ($\vec q\parallel\vec n$): 
\begin{eqnarray}
-2c_V^{\;2}M^{(\nu )00}(q^0,\vec q)&=&m_w^{2}
\left [\frac{1-z}{4}\left(1-\frac{q+q^0}{zq-q^0}\right)\right.
\nonumber \\ \label{M00nup}
&\;&\;\;\;\;\;\;\left. -\frac{1}{2}\frac{q^0}{q}
\left (\ln\left |\frac{zq-q^0}{q-q^0}\right |
-i\pi\Theta (q-q^0)+i\pi\Theta (zq-q^0)\right )\right ] 
\;\;, \\ [1ex]
\label{M0inup} 
M^{(\nu )0i}(q^0,\vec q)&=&M^{(\nu )i0}(q^0, \vec q)\;=\;
\frac{q^0q^i}{q^2}M^{(\nu )00}(q^0,\vec q)
\;\;, \\ [1ex]
\label{MLnup} 
M^{(\nu )}_L(q^0,\vec q)&=&\frac{\omega ^2}{q^2}
M^{(\nu )00}(q^0,\vec q)
\;\;, \\ [1ex]
2c_V^{\;2}M^{(\nu )}_T(q^0,\vec q)&=&
\frac{1}{2}m_w^{2}\frac{\omega^2}{q^2}
\left [\frac{1-z}{2}\left(1
       -\frac{1+z}{2}\frac{q(1-q^2/\omega^2)}{zq-q^0}\right)\right.
\nonumber \\ \label{MTnup} 
&\;&\;\;\;\;\;\; \;\;\;\;\;\;\;\left. -\frac{1}{2}\left (\frac{q^0}{q}
-\frac{q}{q^0}\right )
\left (\ln\left |\frac{zq-q^0}{q-q^0}\right |
-i\pi\Theta (q-q^0)+i\pi\Theta (zq-q^0)\right )\right ] 
,\;\;\;\;\;\; \end{eqnarray}
where $z\equiv\cos\theta_{max}$ and $\omega^2\equiv (q^0)^2$. Indeed, for $z\rightarrow -1$, i.e. without restriction on the opening angle, we recover the formal structure of Eqs.\,(\ref{M00})-(\ref{MT}), while for 
$z\rightarrow 1$ the response function vanishes. 

Furthermore, we observe 
that for a finite opening angle ($1>z>-1$) the Landau damping imaginary 
parts are limited to the region $q>q^0>zq$ and that at the 
resonance frequency $q^0=\Omega_\parallel\equiv zq$ the response function has additional singularities, 
which are absent for $z=-1$. 
In the present case, with $\vec q\parallel\vec n$, the longitudinal as well as the transverse components, 
$M^{(\nu )}_L$ and $M^{(\nu )}_T$ respectively, are affected.   

\underline{{\bf Case II}.} In this case, with 
$\vec q\perp\vec n$, we introduce 
a third unit vector $\vec e$, perpendicular to the other two vectors, 
in order to decompose the momentum vector for the threedimensional 
response function integral,  
\begin{equation}\label{kdecomp1}
\vec k =k(\vec n\cos\theta +\vec q q^{-1}\sin\theta\cos\phi
+\vec e\sin\theta\sin\phi )
\;\;, \end{equation}
with the azimuthal and polar angles $\theta$ and $\phi$, respectively, 
such that $\vec k\cdot\vec q=kq\sin\theta\cos\phi$. The resulting 
angular integrations can all be done analytically in the appropriate
ultrarelativistic limit, either by elementary or 
contour integration techniques. 

Due to the symmetry 
properties of the required integrals, presently it turns out to be useful 
to decompose the spatial part of the response function as follows: 
\begin{equation}\label{decomposition1}
M^{(\nu)ij}(q)\equiv
\left (\delta^{ij}-n^in^j-\frac{q^i q^j}{\vec q^2}\right )M^{(\nu)}_T(q)
+n^in^jM^{(\nu )}_{L_1}(q)
+\frac{q^i q^j}{\vec q^2}M^{(\nu )}_{L_2}(q)
+\frac{n^iq^j+n^jq^i}{|\vec q|}M^{(\nu )}_3(q)
\;\;. \end{equation}
All other terms which could arise vanish identically, since the 
corresponding polar angle integration comprises an odd function.

In order to check the ensuing lengthy 
calculations, we also evaluated independently the 
integrals resulting from the transversality condition mentioned 
after Eq.\,(\ref{response1}), $q_\lambda M^{(\nu)\lambda\rho}(q)=0$, 
as well as from the trace $M_\lambda^{(\nu)\lambda}(q)$. These results  
we compared with what is obtained using the  
calculated components of the response function in the following. 
In fact, this procedure leads to considerable simplifications. 

Then, for $0\leq\theta_{max}\leq\pi /2$, i.e. $0\leq z\leq 1$, we finally obtain these components of the 
{\it neutrino-antineutrino response function} 
($\vec q\perp\vec n$):  
\begin{eqnarray}\label{M00nuperp}
-2c_V^{\;2}M^{(\nu )00}(q^0,\vec q)&=&m_w^{2}
\left [\frac{1\pm 1-z}{4}\mp\frac{z}{4}\frac{q^0}{\sqrt{q_z^2}}
\mp\frac{1}{2}\frac{q^0}{q}
\ln (\frac{q+q^0}{zq+\sqrt{q_z^{\;2}}} )\;
\right ] 
\;\;, \\ [1ex]
\label{M0inuperp}
M^{(\nu )0i}(q^0,\vec q)&=&
M^{(\nu )i0}(q^0,\vec q)\;=\;
\frac{qn^i}{q^0}M^{(\nu )}_3(q^0,\vec q)
+\frac{q^0q^i}{q^2}M^{(\nu )00}(q^0,\vec q)
\;\;, \\ [1ex]
\label{MTnuperp}
2c_V^{\;2}M^{(\nu )}_T(q^0,\vec q)&=&
\frac{1}{2}m_w^{2}\frac{(q^0)^2}{q^2}
\left [\frac{1-z}{2}\mp\frac{1}{2}\left (\frac{q^0}{q}
-\frac{q}{q^0}\right )
\ln (\frac{q+q^0}{zq+\sqrt{q_z^2}} )\;
\right ] 
\;\;, \\ [1ex]
\label{ML1nuperp}
2c_V^{\;2}M^{(\nu )}_{L1}(q^0,\vec q)&=&
\frac{1}{4}m_w^{2}\left [
1-z\pm (\frac{(q^0)^2}{q^2}-1)
\left (
1-z\frac{q^0}{\sqrt{q_z^{\;2}}}-\frac{q^0}{q}
\ln (\frac{q+q^0}{zq+\sqrt{q_z^2}} )\;\right )
\right ]
\;\;, \\ [1ex]
\label{ML2nuperp}
M^{(\nu )}_{L2}(q^0,\vec q)&=&
\frac{(q^0)^2}{q^2}M^{(\nu )00}(q^0,\vec q)
\;\;, \\ [1ex]
\label{M3nuperp}
2c_V^{\;2}M^{(\nu )}_3(q^0,\vec q)&=&
\pm\frac{1}{4}m_w^{2}\frac{q^0}{q}
\left [
\frac{(q^0)^2}{q^2}(1-(q^0)^{-1}\sqrt{q_z^2})
+(\frac{(q^0)^2}{q^2}-1)(1-q^0/\sqrt{q_z^2})\right ]
\;\;, \end{eqnarray}
where we introduced the abbreviation,
\begin{equation}\label{qepsz2}
q_z^2\equiv (q^0)^2-q^2(1-z^2)
\;\;, \end{equation}
with $q\equiv |\vec q|$ and 
$z\equiv\cos\theta_{max}$, as before. Several qualifying remarks are in order here: 
\begin{itemize}
\item For later convenience we did not separate real and 
imaginary parts in Eqs.\,(\ref{M00nuperp})-(\ref{M3nuperp}) which are 
valid for {\it complex} $q^0\equiv\omega +i\gamma$, provided the imaginary 
part here is sufficiently small, $|\gamma |\ll |\omega|$, or infinitesimal. 
\item Either the upper or the lower signs have to chosen consistently in Eqs.\,(\ref{M00nuperp})-(\ref{M3nuperp}) according to the following rules ($0\leq\theta_{max}\leq\pi /2$): 
\begin{eqnarray}\label{signs}
\gamma >0\;\;\mbox{and}\;\;\omega >0\;\Rightarrow\;
\mbox{upper signs}&;&\;\;\;  
\gamma >0\;\;\mbox{and}\;\;\omega <-q\sin\theta_{max}\;\Rightarrow\;
\mbox{lower signs}\;\;;
\nonumber \\ [1ex]
\gamma <0\;\;\mbox{and}\;\;\omega <0\;\Rightarrow\;
\mbox{lower signs}&;&\;\;\;  
\gamma <0\;\;\mbox{and}\;\;\omega >q\sin\theta_{max}\;\Rightarrow\;
\mbox{upper signs}
\;\;. \end{eqnarray} 
They are due to the (angular) contour integrations, which result in different contributions according to the listed rules. 
\end{itemize} 

We do not report the results for $\omega$ in the 
intervals which are excluded in (\ref{signs}), since the azimuthal angle integrations have to be split in this case, yielding even more complicated expressions.  


Obviously, the response function has additional square-root singularities at the resonance frequencies 
$\omega=\Omega_\perp^\pm\equiv\pm  q\sin\theta_{max}$, 
as $\gamma\rightarrow 0$. The transverse component $M^{(\nu )}_T$, 
however, is not affected in the present {\bf Case II} 
($\vec q\perp\vec n$).  
 
This completes the calculation of the response functions for the 
model distributions discussed in the previous subsection. 

\subsubsection{Dispersion Relations, Collective Modes and Instabilities}     
It is useful to begin the study of the dispersion relations following from Eqs.\,(\ref{deltaecurrentfinal})-(\ref{electrondispersion}) with the case 
of the electromagnetically interacting electron-positron plasma, i.e. with  the weak interaction term 
$\propto G_F^{\;2}$ in Eq.\,(\ref{deltaecurrentfinal}) switched off. 
  
Considering separately transverse (`$T$') and longitudinal (`$L$') current fluctuations, i.e. 
$\delta\vec J^{(e)}(q)\perp\vec q$ and 
$\delta\vec J^{(e)}(q)\parallel\vec q$, 
respectively, Eq.\,(\ref{electrondispersion}) yields two 
equations determining the corresponding 
{\it dispersion relations}: 
\begin{eqnarray} \label{Tdispersion}
T:&\;&\left(1-e^2M^{(e)}_T/(q^2+i\epsilon q^0)\right )^3=0
\;\;, \\ [2ex]
\label{Ldispersion}
L:&\;&1-e^2M^{(e)00}/\vec q^2=0
\;\;, \end{eqnarray}
with the plasma response functions of 
Eqs.\,(\ref{M00})-(\ref{MT}), and where     
the decomposition (\ref{decomposition}) of 
the spatial part of the response function is especially
taken into account.  
   
The real solutions with $\omega\equiv |q^0|>q\equiv |\vec q|$ of Eqs.\,(\ref{Tdispersion}) and (\ref{Ldispersion}), respectively, determine the well-known collective {\it transverse and longitudinal plasmon modes} 
\cite{LifshitzP81,Kapusta}. In the 
long-wavelength limit ($\omega\gg q$), for example, 
the explicit solutions are:
\begin{equation}\label{plasmons}
T:\;\;\omega_T^{\;2}(q)=\omega_0^2+\frac{6}{5}q^2
\;\;,\;\;\;\; 
L:\;\;\omega_L^{\;2}(q)=\omega_0^2+\frac{3}{5}q^2
\;\;, \end{equation} 
with the plasma frequency $\omega_0^2\equiv\frac{1}{3}m_D^2$, 
cf. Eq.\,(\ref{Dmass}), 
which characterizes an ultrarelativistic  
neutral plasma. Again this is       
in agreement with the one-loop calculations 
of finite temperature field theory ($T\gg m_e$).
We remark that beyond the present collisionless approximation these modes naturally aquire a finite width \cite{Kapusta}.
   
We now turn to the case of a fully interacting neutrino-antineutrino beam 
impinging on an electron-positron plasma. 
We remind ourselves of the two limiting cases introduced in the 
preceding section concerning the orientation of the wave vector 
$\vec q$ with respect to the outward normal vector $\vec n$, i.e.  
$\vec q\parallel\vec n$ ({\bf Case I}) and $\vec q\perp\vec n$ 
({\bf Case II}). 
In both cases, we concentrate 
on the interesting possibility that the weak interaction 
term might become comparable to the purely electromagnetic term $\propto e^2$ in Eq.\,(\ref{electrondispersion}). Due to the intrinsic smallness 
of the weak coupling constant this may happen only, when the neutrino-antineutrino response functions become large, close to the 
singularities found in Eqs.\,(\ref{M00nup})-(\ref{MTnup}) or in 
Eqs.\,(\ref{M00nuperp})-(\ref{M3nuperp}). Otherwise the neutrino 
effects can be treated as small perturbations of previous 
electron-positron plasma results, as we shall see. 
  
\underline{{\bf Case I}.} We observe here that the neutrino-antineutrino response function obtained in Eqs.\,(\ref{M00nup})-(\ref{MTnup}) has the same 
tensor structure as the electron-positron one of   
Eqs.\,(\ref{M00})-(\ref{MT}). Considering the product of the two appearing in Eq.\,(\ref{electrondispersion}), 
${\cal C}^{\alpha\beta}\equiv M^{(e)\alpha\gamma}M^{(\nu )\;\beta}_{\;\gamma}$, we obtain: 
\begin{eqnarray}\label{product1}
{\cal C}^{00}&=&(1-\frac{\omega^2}{q^2})M^{(e)00}M^{(\nu )00}
\;\;, \\
\label{product2}
{\cal C}^{0i}&=&{\cal C}^{i0}=\frac{q^0q^i}{q^2}{\cal C}^{00}
\;\;, \\
\label{product3}
{\cal C}^{ij}&=&-\left (\delta^{ij}-\frac{q^iq^j}{q^2}\right )
M^{(e)}_TM^{(\nu )}_T+\frac{q^iq^j}{q^2}\frac{\omega^2}{q^2}{\cal C}^{00} 
\;\;, \end{eqnarray}
with $\omega\equiv |q^0|$ and $q\equiv |\vec q|$. 
Therefore, we may still distinguish 
transverse ($T$) and longitudinal ($L$) current fluctuations which do not mix, similarly to the case of a purely electromagnetic plasma. 

In analogy to Eqs.\,(\ref{Tdispersion}) and (\ref{Ldispersion}), we 
thus obtain from Eq.\,(\ref{electrondispersion}) two equations which now determine the {\it neutrino `beam' electron-positron plasma dispersion relations} ($\vec q\parallel\vec n$):   
\begin{eqnarray}\label{Tnudispersion}
&T:&\left (1-[\frac{e^2}{(q^0)^2-q^2}
+2c_V^{\;2}G_F^{\;2}M^{(\nu )}_T]M^{(e)}_T\right )^3=0
\;\;, \\[2ex]
\label{Lnudispersion}
&L:&1-[\frac{e^2}{q^2}
+2c_V^{\;2}G_F^{\;2}(1-\frac{(q^0)^2}{q^2})^2
M^{(\nu )00}]M^{(e)00}=0
\;\;, \end{eqnarray}
with $q\equiv |\vec q|$. 

In the long-wavelength limit ($\omega\gg q$) and to lowest order in 
$G_F^{\;2}$ we obtain, for example, from Eq.\,(\ref{Tnudispersion}) 
the equation: 
\begin{eqnarray} 
0&=&\omega^2-q^2-\omega_0^2(1+\frac{1}{5}(\frac{q}{\omega})^2+\dots\;)  
\nonumber \\ \label{Tnudispersion1} 
&\;& 
-\omega_0^2a^2
(\frac{4(1-z)+z(1-z^2)}{36}-\frac{(1-z^2)^2}{24}\frac{q}{\omega}
+\frac{4(1-z)-3z(1-z^2)^2}{60}
(\frac{q}{\omega})^2+\dots\;)
\;\;, \end{eqnarray} 
with $\omega\equiv |q^0|$, $z\equiv\cos\theta_{max}$, and 
where we indicated the neglected higher order terms in $q/\omega$. 
We introduced the dimensionless constant: 
\begin{equation}\label{a2}
a^2\equiv\frac{1}{2e^2}G_F^{\;2}m_w^2m_D^2  
\;\;,  \end{equation}
which governs the strength of the neutrino effects.  
The solution of Eq.\,(\ref{Tnudispersion1}) describes the {\it transverse plasmon in a neutrino `beam' electron-positron plasma}. 

However, as we anticipated, the smallness of the 
weak coupling constant makes the influence of the neutrino terms 
completely negligible here. Considering Type\,II supernova 
conditions and setting $T_e\approx 1$\,MeV and $T_\nu\approx 10$\,MeV, we find that $a^2\approx 10^{-22}$. Omitting the neutrino 
contribution and solving reproduces the first of Eqs.\,(\ref{plasmons}).    

A similar analysis, i.e. for $\omega >q$, applies to Eq.\,(\ref{Lnudispersion}) which   
the {\it longitudinal plasmon in a neutrino `beam' electron-positron plasma}. Again the neutrinos have a  
negligible effect under supernova conditions. 
 
We now consider the   
dispersion relations implicit in Eqs.\,(\ref{Tnudispersion}) and (\ref{Lnudispersion}) close to the 
resonance frequency, i.e. $\omega\approx\Omega_\parallel\equiv zq$, which lies in the electron-positron 
Landau damping regime with $0<\omega <q$, considering $0<z<1$ from  
now on ($0<\theta_{max}<\pi /2$).\footnote{For the case of an  
ultrarelativistic pure electron-positron plasma in equilibrium it can 
be shown that no solution, for example, of the dispersion equation (\ref{Ldispersion}) exists with $0<\omega <q$.}    

In this case, we expect the
frequency $q^0$, and correspondingly $\omega$, to aquire a {\it finite} 
imaginary part, instead of the infinitesimal $i\epsilon$ representing 
the retarded boundary condition \cite{LifshitzP81}, cf. 
Eqs.\,(\ref{deltaeVlasovSol}),\,(\ref{deltanVlasovSol}) 
or (\ref{response1}). Therefore, 
replacing $q^0+i\epsilon\longrightarrow\omega +i\gamma$, the `Landau 
logarithms' and imaginary parts of the calculated response functions 
have to be reconsidered. We rewrite $\omega\equiv zq+\xi$, 
anticipating that $\xi\ll qz$, and will use:
\begin{equation}\label{Llog}
\ln\frac{\omega +q+i\gamma}{\omega -q+i\gamma}
=\ln\frac{1+z}{1-z}-i\pi\mbox{Sign}\gamma
+\ln (1+\frac{\xi +i\gamma}{q(1+z)})
-\ln (1-\frac{\xi +i\gamma}{q(1-z)}) 
\;\;, \end{equation}
where Sign$\gamma\equiv\gamma /|\gamma |$\,. This  
is most appropriate for small $\xi$ and $\gamma$, 
reproducing the usual result for $\gamma\rightarrow 0^+$. 
 
Specifically, we reconsider Eq.\,(\ref{Lnudispersion}) and take only 
the dominant singular term $\propto (zq-q^0)^{-1}$ in $M^{(\nu )00}$ 
into account, cf. Eq.\,(\ref{M00nup}). Thus we obtain more explicitly: 
\begin{equation}\label{Lnusing} 
1+\left[\frac{m_D^2}{q^2}+a^2\frac{1-z}{2}
(1-\frac{(\omega +i\gamma )^2}{q^2})^2
\frac{q+\omega +i\gamma}{zq-\omega -i\gamma}\right]
\left[1-\frac{1}{2}\frac{\omega +i\gamma}{q}\ln\frac{\omega +q+i\gamma}
{\omega -q+i\gamma}\right]=0 
\;\;, \end{equation}
with $\omega\equiv zq+\xi$. We recall that $a^2\ll 1$.  

It is easy to see that for a solution with $\omega\approx zq$ the 
term $\propto a^2$ has to behave qualitatively such that (at least)   
$(\xi ,\gamma )/q\sim a^2q^2/m_D^2\ll 1$, particularly in the 
long-wavelength limit with 
$q^2/m_D^2\ll 1$. Consequently, using Eq.\,(\ref{Llog}), 
we expand Eq.\,(\ref{Lnusing}) up to second order in $\xi /q$ or 
$\gamma /q$. Separating real and imaginary parts, it is straightforward 
to solve the resulting equations. We obtain: 
\begin{eqnarray} \label{Lxi}
\frac{\xi}{q}&=&\frac{1}{2}a^2(q/m_D)^2(1-z^2)^3
\frac{(\pi z/2)^2+f^2(z)+f(z)(q/m_D)^2}
{(\pi z/2)^2+[f(z)+(q/m_D)^2]^2}
\;\;, \\ [1ex]
\label{Lgamma}
\frac{|\gamma |}{q}&=&\frac{\pi}{2}a^2(q/m_D)^4\frac{(1-z^2)^3}
{(\pi z/2)^2+[f(z)+(q/m_D)^2]^2}
\;\;, \end{eqnarray}
neglecting higher order in $a^2$ corrections and defining: 
\begin{equation}\label{f}
f(z)\equiv 1-\frac{1}{2}z\ln\frac{1+z}{1-z} 
\;\;. \end{equation}
These results are consistent with the applied expansions, noting that 
$\xi\propto G_F^{\;2}/e^2$ and $\gamma\propto G_F^{\;2}/e^4$,  
particularly in the long-wavelength limit.   
  
Recalling $\omega\equiv zq+\xi$, we thus obtain a pair of 
{\it longitudinal pharon modes} (`Type\,I', i.e. for 
$\vec q\parallel\vec n$), with 
the real part of the dispersion relation in the long-wavelength 
limit given by: 
\begin{equation}\label{Lomega} 
\omega (q)=zq+\frac{1}{4e^2}G_F^{\;2}m_w^2(1-z^2)^3q^3
\;\;, \end{equation}
one, a growing and the other, a decaying mode, depending on  
the sign of $\gamma$. 
  
Analogously, we analyze the transverse dispersion relation to be 
calculated from Eq.\,(\ref{Tnudispersion}) for $\omega <q$.  
In this case, we find a pair of {\it transverse pharon modes} 
(Type\,I) with:
\begin{eqnarray}\label{Txi} 
\frac{\xi}{q}&=&-\frac{1}{4}a^2(q/m_D)^2(1-z^2)^3
\frac{(\pi /2)^2(1-z^2)+(z^2/(1-z^2))g^2(z)+2g(z)(q/m_D)^2}
{(\pi z/2)^2+[(z^2/(1-z^2))g(z)+2(q/m_D)^2]^2}
\;\;, \\ [1ex] 
\label{Tgamma}
\frac{|\gamma |}{q}&=&\frac{\pi}{4}a^2(q/m_D)^4\frac{z^{-1}(1-z^2)^3}
{(\pi z/2)^2+[(z^2/(1-z^2))g(z)+2(q/m_D)^2]^2}
\;\;, \end{eqnarray}
and where: 
\begin{equation}\label{g}
g(z)\equiv 1-\frac{1}{2}(z-\frac{1}{z})\ln\frac{1+z}{1-z}
\;\;. \end{equation}
The corresponding real part of the dispersion relation in the 
long-wavelength limit is: 
\begin{equation}\label{Tomega} 
\omega (q)=zq-\frac{1}{8e^2}G_F^{\;2}m_w^2(1-z^2)^3q^3
\frac{(\pi /2)^2(1-z^2)+(z^2/(1-z^2))g^2(z)}
{(\pi z/2)^2+(z^2/(1-z^2))^2g^2(z)}
\;\;, \end{equation}
with an interesting negative sign in front of 
the second term. 
We observe that the transverse pharons are quite sensitive 
to the geometry parameter $z$. 

In particular, in the limit 
$z\rightarrow 0$, corresponding to a maximally fanned-out `beam' with   
$\theta_{max}\rightarrow\pi /2$, the `damping constant' $\gamma (q)$ 
diverges. In this limit the expansions leading to Eqs.\,(\ref{Txi}) and 
(\ref{Tgamma}) clearly break down. This can be studied 
in more detail, starting again with Eq.\,(\ref{Tnudispersion}) and implementing $q^0=\xi +i\gamma$, with $\xi ,\gamma\ll q$. However, it 
leads to a nonpropagating mode with frequency of the same small order of 
magnitude as the damping constant, which is physically irrelevant to our 
study.     

However, under Type\,II supernova conditions, with 
$a^2\approx 10^{-22}$, and recalling that we have 
$1-z^2=\sin^2\theta_{max}=(R/r)^2$, in terms of the radius $R$ of the 
neutrino sphere and the distance $r$ of the electron-positron plasma from 
its center, a typical value may be $(R/r)^2\approx 0.5$ for $R\approx 30$\,km \cite{RECENT1}. Then, for $q<m_D$, the above calculations are  
accurate and we may roughly estimate, for example, the transverse 
pharon damping constant,   
$\gamma\approx 10^{-2}a^2(q/m_D)^4q$. For a
pharon wavelength corresponding to $q\approx m_D/2$ and an electron 
temperature $T_e\approx 1$\,MeV, this yields a   
growth/decay length (one e-folding) on the order of $10^9-10^{11}$\,km. 
A one-percent increase of the collective mode amplitude squared, 
i.e. of its energy, means it would have to run through more than 
$10^6$\,km of plasma, which is simply 
not there. The longitudinal mode behaves similarly.   

Clearly, the above estimates are crude and could be improved by 
folding the results with the appropriate distributions, depending on 
the distance from the supernova core (and time). However, in view of 
the intrinsic weakness of the instabilities, we conclude that it is unlikely that long-wavelength Type\,I ($\vec q\parallel\vec n$) pharon modes play an   important role in the outward energy transport processes in Type\,II supernovae. 

Considering the strong momentum dependence 
of the calculated damping constants, Eqs.\,(\ref{Lgamma}) and 
(\ref{Tgamma}), however, the question is raised, whether, at  
shorter wavelengths, corresponding collective modes   
could become important instead. As we will discuss in the 
following subsection in more detail, in this limit, the presently employed semiclassical transport theory breaks down, necessitating further study.   

\underline{{\bf Case II}.} We recall that here we have $\vec q\perp\vec n$ 
and proceed as before. However, the product of the two response matrices appearing in Eq.\,(\ref{electrondispersion}),   
${\cal C}^{\alpha\beta}\equiv M^{(e)\alpha\gamma}M^{(\nu )\;\beta}_{\;\gamma}$, has to be recalculated. Taking the different 
tensor structure of $M^{(\nu )}$, according to 
Eqs.\,(\ref{M00nuperp})-(\ref{M3nuperp}), into 
account, we obtain: 
\begin{eqnarray}\label{product10}
{\cal C}^{00}&=&(1-\frac{(q^0)^2}{q^2})M^{(e)00}M^{(\nu )00}
\;\;, \\
\label{product20}
{\cal C}^{0i}&=&\frac{q^0q^i}{q^2}{\cal C}^{00}
+n^i(\frac{q}{q^0}-\frac{q^0}{q})M^{(e)00}M_3^{(\nu )}
\;\;, \\
\label{product30}
{\cal C}^{i0}&=&\frac{q^0q^i}{q^2}{\cal C}^{00}
-\frac{qn^i}{q^0}M_T^{(e)}M_3^{(\nu )}
\;\;, \\ 
{\cal C}^{ij}&=&
\frac{q^iq^j}{q^2}\frac{(q^0)^2}{q^2}{\cal C}^{00} 
+\frac{q^in^j}{q}(1-\frac{(q^0)^2}{q^2})M^{(e)00}M_3^{(\nu )}
\nonumber \\
\label{product40}
&\;&-\left (
(\delta^{ij}-n^in^j-\frac{q^iq^j}{q^2})M^{(\nu )}_T
+n^in^jM^{(\nu )}_{L1}
+\frac{n^iq^j}{q}M_3^{(\nu )}\right )
M^{(e)}_T
\;\;, \end{eqnarray}
with $q\equiv |\vec q|$.   
We observe that ${\cal C}^{\mu\nu}\neq {\cal C}^{\nu\mu}$. 
 
In the present case, we consider again two different kinds of current 
fluctuations, when evaluating Eq.\,(\ref{electrondispersion}): 
$\delta\vec J^{(e)}(q)\perp\vec q,\vec n$ (`$Out$'), i.e. 
fluctuations which are 
perpendicular to the plane spanned by $\vec q$ and $\vec n$, 
and fluctuations with $\delta\vec J^{(e)}(q)$ in this plane (`$In$').  
Thus we obtain the following two equations which determine the 
{\it neutrino `beam' electron-positron plasma dispersion relations} 
($\vec q\perp\vec n$):   
\begin{eqnarray}\label{Tnudispersion0}
&Out:&\left (1-[\frac{e^2}{(q^0)^2-q^2}
+2c_V^{\;2}G_F^{\;2}M^{(\nu )}_T]M^{(e)}_T\right )^3=0
\;\;, \\[2ex]
&In:&
\left (1-2c_V^{\;2}G_F^{\;2}M^{(\nu )}_{L1}M^{(e)}_T\right )
\left (1-[\frac{e^2}{q^2}
+2c_V^{\;2}G_F^{\;2}(1-\frac{(q^0)^2}{q^2})^2
M^{(\nu )00}]M^{(e)00}\right )
\nonumber \\
\label{Lnudispersion0}
&\;&+(2c_V^{\;2}G_F^{\;2})^2
(2-\frac{q^2}{(q^0)^2}-\frac{(q^0)^2}{q^2})
(M^{(\nu )}_3)^2M^{(e)}_TM^{(e)00}
=0
\;\;, \end{eqnarray}
with $q^0\equiv\omega +i\gamma$. 
We observe that Eq.\,(\ref{Tnudispersion0}) has the same formal 
structure as Eq.\,(\ref{Tnudispersion}) before. Furthermore, if we set $G_F^{\;2}$ to zero, these equations reproduce the transverse and longitudinal electron-positron plasmon dispersion equations (\ref{Tdispersion}) and (\ref{Ldispersion}).    
  
Guided by our analysis of the transverse plasmon dispersion relation 
under neutrino flux for $\vec q\parallel\vec n$, Eqs.\,(\ref{Tnudispersion})-(\ref{a2}), we expect  
only a negligible perturbative influence    
of the neutrino interactions in Eq.\,(\ref{Tnudispersion0}), since 
they are again suppressed by the factor $a^2\approx 10^{-22}$. In particular, the present structure of $M_T^{(\nu )}$ is a smoothly  
deformed version of $M_T^{(e)}$, compare Eqs.\,(\ref{MT}) and (\ref{MTnuperp}), with no additional singularity. Thus, the 
corresponding collective 
mode shows no particularly interesting behavior and presents a 
{\it second kind of perturbatively deformed transverse plasmon}.  
  
Finally, we consider Eq.\,(\ref{Lnudispersion0}), describing a 
geometry which resembles the one where two-stream instabilities 
arise in other plasmas \cite{Weibel59,LifshitzP81,Stan96}. We attempt 
to find pharon type solutions in the present case as well. For this 
purpose, we take into account the leading root-singular terms, which contribute here from 
Eqs.\,(\ref{M00nuperp}), (\ref{ML1nuperp}), and (\ref{M3nuperp}).  
The singularities occur at the resonance frequencies  
$\omega=\Omega_\perp^\pm\equiv\pm qs$, as   
$q_z^2\equiv (q^0)^2-q^2(1-z^2)\rightarrow 0$, where 
$q\equiv |\vec q|$ and $z\equiv\cos\theta_{max}$.   

We concentrate on the positive frequency solutions of 
the dispersion equation (\ref{Lnudispersion0}) and are particularly 
interested in those with a positive 
imaginary part, which grow exponentially in time. 
Therefore, we consider $q^0\equiv\omega +i\gamma$, with 
$\omega\approx qs>0$, defining $s\equiv\sin\theta_{max}$. Then,  
Eq.\,(\ref{Lnudispersion0}) assumes a slightly simpler form: 
\begin{equation}\label{Lnudispersion1}  
0=1+m_q^2\tilde M^{(e)00}
+
a^2\frac{z}{2}\frac{q^0}{\sqrt{q_z^2}}
[\frac{(q^0)^2}{q^2}-1]
\left ([\frac{(q^0)^2}{q^2}-1]\tilde M^{(e)00}+
\frac{1}{2}\tilde M^{(e)}_T [1+m_q^2\tilde M^{(e)00}]\right )
+O(a^4)
\;\;, \end{equation} 
with the dimensionless effective coupling constant $a^2$, Eq.\,(\ref{a2}).  
The terms of $O(a^4)$ will be neglected in the following, since their  
singularities cancel. Furthermore, we conveniently define:
\begin{equation}\label{abbr}
m_q^2\equiv m_D^2/q^2\;\;,\;\;\;
\tilde M^{(e)00}\equiv -e^2M^{(e)00}/m_D^2\;\;,\;\;\;
\tilde M^{(e)}_T\equiv 2e^2M^{(e)}_T/m_D^2
\;\;, \end{equation}
cf. Eqs.\,(\ref{M00}) and (\ref{MT}). Recalling the smallness of 
$a^2$, it is obvious that any interesting solution must arise  
close the resonance frequency $\Omega_\perp^+=qs$ ($0<s<1$).   
  
Setting $\omega\equiv qs+\xi$ and assuming $\xi ,\gamma\ll qs$, it is 
useful to expand Eq.\,(\ref{Lnudispersion1}) in terms of the small 
complex quantity $\kappa\equiv (\xi +i\gamma )/(2qs)$. Here we make use 
of Eq.\,(\ref{Llog}) once more, in order to expand the Landau logarithms.
Then, expanding to leading order in $\kappa$, it is straightforward to arrive at the `formal solution': 
\begin{equation}\label{kappa}
\xi +i\gamma =qa^4\frac{s(1-s^2)^3}
{128[(1+m_q^2f(s))^2+(\pi s/2)^2]^2}
\left (h(s)
[1+m_q^2(f(s)-(i\pi s/2)\mbox{Sign}\gamma )]
\right )^2 
\;\;, \end{equation}
where: 
\begin{eqnarray}\label{hofs}
h(s)&\equiv&2s^2g(s)+i\pi s(s^2-1)\mbox{Sign}\gamma 
+4(s^2-1)(f(s)+(i\pi s/2)\mbox{Sign}\gamma )
\nonumber \\ [1ex]
&\;&+m_q^2(f(s)+(i\pi s/2)\mbox{Sign}\gamma )
(2s^2g(s)+i\pi s(s^2-1)\mbox{Sign}\gamma )
\;\;. \end{eqnarray}
The functions $f$ and $g$ were defined in Eqs.\,(\ref{f}) and 
(\ref{g}). 
The appearance of $\mbox{Sign}\gamma$ on the right-hand side 
restricts the possibility of 
an explicit solution. 
  
  
After some algebra, one obtains a criterion for a solution to 
exist in the relevant regime ($0<s<1$):  
\begin{equation}\label{crit}
\left (-1+s^2[1+m_q^2g(s)/2]\right )
\left (\pi^2m_q^2s^2+4f(s)[1+m_q^2f(s)]\right )
+2s^2g(s)[1+m_q^2f(s)]<0
\;\;, \end{equation} 
which in the long-wavelength limit ($m_q^2>1$) can only be fullfilled for sufficiently small $s$, i.e. sufficiently small opening angle of the neutrino momentum distribution. Clearly, taking only the leading terms 
in this limit into account, no solution exists. On the other hand, for $m_D/q=2$, for example, the solvability criterion requires $s<0.426$, 
corresponding to $\theta_{max}\approx 25^o$.   
 
It is obvious, however, from Eq.\,(\ref{kappa}) that any 
{\it pharon} (`Type\,II', i.e. for $\vec q\perp\vec n$) solution here 
will have $\xi ,\gamma\propto a^4\propto G_F^{\;4}$, with no particular  factors especially enhancing the damping constant.  
We refrain from giving explicitly the not very illuminating lengthy expressions. 

Instead, we conclude that in the supernova environment the growth rate 
of the presently studied Type\,II pharons is suppressed by an extra factor 
of $a^2\approx 10^{-22}$ , as compared to Type\,I. Consequently, these 
modes do not contribute at all in this case.
    
\subsubsection{Discussion}
The detailed calculations in the previous subsections are based on the stationary electron and neutrino distribution functions, $f_S^{(e)}(k)$ and $f_S^{(\nu )}(k)$ of 
Eqs.\,(\ref{Te0density}) and (\ref{Tn0density}), respectively. 
While $f_S^{(e)}(k)$ is adequate for the supernova scenario discussed here, the question arises as to whether the 
sharp {\it azimuthal angle cut-off}, present in $f_S^{(\nu )}(k)$, may not cause spurious effects or invalidate our semiclassical transport approach. 
     
In fact, the {\it semiclassical approximation} of Section\,2.2 is based on the expansion of the full quantum transport equations in powers of $\hbar$, appearing especially in the 
dimensionless combination $\hbar\partial_x\cdot\partial_p$, cf.  Eqs.\,(\ref{transport}) and (\ref{K})-(\ref{Pi}) in Section\,2.1. 
Therefore, a cut-off on a spacelike momentum coordinate, corresponding to the angle $\theta_{max}$ between three-momentum and outward normal direction, may produce large higher order corrections. These are controlled, however, in the {\it long-wavelength limit}. 
We recall that in the derivation of the linear response theory in Section 3.1, the space-time gradients $\partial /\partial x^\mu$ become the four-momenta $q_\mu$, beginning with Eq.\,(\ref{deltaeVlasov}). 
In the long-wavelength limit, it is generally required that $q$, which probes the spatial inhomogeneity of 
the (stationary) system, be small compared to the relevant momentum (gradient) scales, i.e. the temperatures $T_e,T_\nu$\,.
Otherwise, the response functions, see Eq.\,(\ref{response1}), would inherit neglected higher order terms in $q^\mu \partial_k^{\;\nu}$, which correspond to going from Eq.\,(\ref{transport}) to Eq.\,(\ref{Vlasov}). 

A truly microscopic transport calculation of the neutrino distribution, as they are released from the neutrino sphere, is an interesting topic for future work \cite{Cooperstein88,RECENT1}. We wish to conclude by illustrating the modifications resulting from a more realistic smooth cut-off neutrino momentum distribution. 
 
For example, we consider the azimuthal angle integral which contributes the singular term $\propto (q+q^0)/(zq-q^0)$ to the neutrino response function $M^{(\nu )00}$, Eq.\,(\ref{M00nup}), which in turn is essential for the longitudinal Type\,I pharon originating in Eq.\,(\ref{Lnudispersion}). Following the radial momentum and polar angle integrations, one encounters the integral: \begin{equation}\label{I}
I\equiv q^2\int_{-1}^1dz\;(q^0+i\epsilon -|\vec q|z)^{-2}F(z) \;\;, \end{equation} 
where we replaced the previous sharp cut-off, i.e. $\Theta (z-z_m)$, by the smooth function $F$, \begin{equation}\label{F} 
F(z)\equiv N_+^{-1}
(\arctan [\sqrt{\alpha}(z-z_m)]+\arctan [\sqrt{\alpha}(1+z_m)]) \;\;, \end{equation} 
where $z_m\equiv\cos\theta_{max}$, and we have introduced the convenient abbreviations:
\begin{equation}\label{N}
N_\pm\equiv\arctan [\sqrt{\alpha}(1-z_m)]\pm\arctan [\sqrt{\alpha}(1+z_m)] \;\;. \end{equation}
Thus, we have $F(-1)=0$, $F(1)=1$ and the Lorentzian derivative, \begin{equation}\label{Fderiv} 
F'(z)=\frac{\sqrt{\alpha}N_+^{-1}}
{1+\alpha (z-z_m)^2}
\;\;. \end{equation}
The parameter $\alpha >0$ determines the steepness of the sigmoid cut-off, which ultimately should be related to the emission characteristics of the neutrino sphere \cite{RECENT1}. 
 
Then, after a partial integration, we obtain an integral with three 
complex simple poles,  
\begin{equation}\label{Ipartial}
I=\frac{q^0+|\vec q|}{|\vec q|}+
\frac{q^2}{|\vec q|^2}\int_{-1}^1dz\;
\frac{F'(z)}{z-(q^0+i\epsilon )/|\vec q|} \;\;, \end{equation}
which can be solved analytically. 
The final result is: 
\begin{eqnarray}\label{Ifinal}
I&=&1+\frac{q^0}{|\vec q|}
+\frac{q^2\left (\sqrt{\alpha}N_+\right )^{-1}} {(q^0-|\vec q|z_m)^2+|\vec q|^2/\alpha} [\sqrt{\alpha}N_+(z_m-\frac{q^0}{|\vec q|}) +\ln\frac{q^0-|\vec q|}{q^0+|\vec q|}
+\frac{1}{2}\ln\frac{1+\alpha (1+z_m)^2}{1+\alpha (1-z_m)^2}] \nonumber \\ [2ex]
&\;&\stackrel{a\rightarrow\infty}{\longrightarrow}\; (1-z_m)\frac{q^0+|\vec q|}{q^0-|\vec q|z_m} \;\;. \end{eqnarray}
Thus the previous result is recovered in the sharp cut-off limit.
       
We observe that the previous {\it resonance pole} at $q^0=\Omega_\parallel\equiv|\vec q|z_m$ is moved symmetrically off the real axis. However, closer inspection shows that the above result is {\it not} singular any more as $q^0\rightarrow |\vec q|(z_m\pm i/\sqrt\alpha )$. We recalculated the complete $M^{(\nu )00}$-component of the neutrino response function with the smooth cut-off function $F(z)$ and found a corresponding result. 
 
Therefore, in accordance with intuitive expectation, a physically more appropriate moderate smoothing of the angular dependence of the neutrino momentum distribution turns the pharon resonance poles into resonances with a finite width, which decreases with increasing sharpness of the cut-off (i.e. as $\alpha\rightarrow\infty$). 
Most likely, this implies that realistic growth rates for the unstable modes will be lower than estimated optimistically in Section 3.2.3\,.

\section{Conclusions}
In this work, we studied the collective modes which were earlier conjectured to produce anomalously large instabilities in the course of the interaction of the neutrino `beam' with the plasma sphere in Type\,II supernovae \cite{Bingham94}. 
 
For this purpose, we derived the semiclassical transport theory based on the Dirac field equations for neutrinos and electrons, which are coupled according to the Standard Model. 
Our results also allow for the handling of situations with strong spin-polarizing 
magnetic fields, which we did not consider here. We derived a related linear response theory and applied it in a detailed supernova scenario, adapted from Bethe's review \cite{RECENT1}. 

We studied, in particular, the modifications of the usual transverse and longitudinal {\it plasmons} of an electron-positron plasma, which are caused by a high-power beam-like flux of neutrinos, such as the almost radially outward streaming neutrinos, which are released from the neutrino sphere surrounding the supernova core. 
In the collisionless approximation, we found only a very weak perturbative effect and no induced imaginary part of the dispersion relation. This is due to the suppression of all neutrino effects by the small dimensionless effective coupling constant, $$a^2\equiv\frac{1}{2e^2}G_F^{\;2}m_w^2m_D^2\approx 10^{-22} \;\;, $$ 
under typical Type\,II supernova conditions.   
    
However, we found new types of growing, as well as damped collective modes, the {\it pharons}, which are characterized by an essentially linear dispersion relation $\omega (q)/q\approx const$ in the long-wavelength limit. Their characteristic properties depend on the relative orientation of the neutrino beam, collective mode propagation and electric current fluctuation directions (the {\bf Cases I} and {\bf II} are studied in Sections 3.2.2 and 3.2.3).  
They partially overcome the discussed suppression, since a resonance pole arises in the dispersion equations in the region with $\omega (q)<q$. In this region, one finds in an ordinary electron-positron plasma strongly    Landau damped modes or, in the ultrarelativistic limit, no (plasmon) modes at all. 

As we estimated roughly in Section 3.2.3, although they partially overcome the discussed suppression, the Type\,I pharon growth rates are still about four, and likely more, orders of magnitude too small to make an impact on supernova evolution. The Type\,II pharon growth rate is even more suppressed; it is proportional to $G_F^{\;4}$, like a purely weak interaction cross section. 
Clearly, we have seen that for the electro-weak interaction to be relevant, the effective particle densities have to be sufficiently high, since one has to overcome the suppression expressed by the smallness of $a^2$, given above, or its equivalent for more massive particles. 
We remark that changing the neutrino and electron temperatures, as compared to the typical supernova case discussed after Eq.\,(\ref{Tomega}), for example,  
while keeping the `wavelength' $m_D/q$ fixed, the Type\,I pharon damping constants grow and the corresponding e-folding lengths drop $\propto T_\nu^{-2}T_e^{-3}$. 

At this point it is worthwhile remarking that more realistic calculations also have to take the collisonal damping into account. This can be incorporated in our approach in the relaxation time approximation in future applications \cite{LifshitzP81}.   

Finally, we point out that pharon type modes should occur in other situations with a current-current type interaction under two-stream conditions. The anisotropic momentum distribution characteristic of a `beam' with limited (momentum) opening angle appears to cause the resonance effect exciting these modes by impact on an isotropic plasma. -- We did not study here a spatially limited beam or jet, which causes quite different `hydrodynamic' instabilities. 

Pharon modes may perhaps be fed effectively by the still unknown central engine powering gamma ray bursts, for example, see Refs.\,\cite{GRB}.
There is growing evidence for even truly jet-like processes in the gamma ray burst phenomenon. Furthermore, allowing for other than electro-weak interactions, such modes possibly come into play in the ultimate evaporation of primordial black holes \cite{Kapusta99}.

To summarize, we conclude that the intrinsic weakness of the neutrino caused collective effects, 
related to the large asymmetry of the electromagnetic and weak coupling strengths, makes it rather unlikely that they play a role in the neutrino
energy deposition in the supernova plasma sphere. However, the new pharon type instabilities may be quite relevant in two-stream situations occuring in other astrophysical systems.     

\subsection*{Acknowledgements}
H.T.E wishes to thank U.\,Heinz, St.\,Mr\'{o}wczy\'{n}ski, and L.\,O.\,Silva for  correspondence. -- H.T.E and T.K. were supported in part by PRONEX (No. 
41.96.0886.00), R.O. would like to thank PRONEX/ FINEP (No. 41.96.0908.00) for partial support, and all three of us acknowledge partial support 
by CNPq-Brasil. 

\section*{Appendix}

Here we consider in more detail the structure of the neutrino-antineutrino spinor Wigner function.  
The general results of Eqs.\,(\ref{spindecomp})-(\ref{antisymt}) can be further specialized for 
the case of the Standard Model $\nu_L\bar\nu_R$-system in the massless limit. -- We follow the 
notation of Ref.\,\cite{HalzenMartin} in this Appendix.   
  
In the main part of the paper we use the Dirac-Pauli representation of the $\gamma$-matrices, 
which are defined by the anticommutation relations $\{\gamma^\mu ,\gamma^\nu \}=2g^{\mu\nu}$, 
i.e. explicitly:
\begin{equation}\label{gammaD}
\gamma^0_D=\left (\begin{array}{cc}1 & 0 \\ 
                                   0 & -1 \end{array}\right )\;\;,\;\;\; 
\gamma^i_D=\left (\begin{array}{cc}0 & \sigma_i \\ 
                                   -\sigma^i & 0 \end{array}\right )\;\;;\;\;\; 
\gamma^5_D=\left (\begin{array}{cc}0 & 1 \\ 
                                   1 & 0 \end{array}\right )
\;\;, \end{equation}
where all entries are $2\times 2$-matrices themselves; in particular,  
$\sigma^i,i=1,2,3$ denote the standard Pauli matrices. We listed also the chirality operator 
$\gamma^5\equiv i\gamma^0\gamma^1\gamma^2\gamma^3$, which anticommutes with all $\gamma^\mu$. 
In the following we will conveniently begin with the chiral Weyl representation:             
\begin{equation}\label{gammaW}
\gamma^0_W=\left (\begin{array}{cc}0 & 1 \\ 
                                   1 & 0 \end{array}\right )\;\;,\;\;\; 
\gamma^i_W=\left (\begin{array}{cc}0 & \sigma_i \\ 
                                   -\sigma^i & 0 \end{array}\right )\;\;;\;\;\; 
\gamma^5_W=\left (\begin{array}{cc}-1 & 0 \\ 
                                   0 & 1 \end{array}\right )
\;\;. \end{equation}
The subscripts $D,W$ presently serve to distinguish the representations, which are related by:  
\begin{equation}\label{Utransf} 
\gamma^\mu_D=U_W\gamma^\mu_WU^\dag_W\;\;,\;\;\;U_W\equiv\frac{1}{\surd 2}(1-\gamma^5_W\gamma^0_W)
\;\; \end{equation} 
i.e. a simple unitary transformation.

In the Weyl representation the stationary Dirac equation separates into a pair of  
two-component Weyl equations: 
\begin{eqnarray}\label{WeylEq1}
E\chi (\vec p)=-\vec\sigma\cdot\vec p\;\chi (\vec p)
\;\;, \\ [1ex] \label{WeylEq2}
E\phi (\vec p)=+\vec\sigma\cdot\vec p\;\phi (\vec p)
\;\;, \end{eqnarray} 
the first of which describes the physical $\nu_l\bar\nu_R$-system, as we shall see, 
while the second one is discarded in the Standard Model. Using $\{\sigma^i,\sigma^j\}=2\delta^{ij}$, 
one verifies that $E^2=|\vec p|^2$, in both cases. Concentrating on Eq.\,(\ref{WeylEq1}), 
we consider first the positive energy solution with four-momentum $p^\mu_+ =(E>0,\vec p)$, 
which obeys ($\hat p\equiv\vec p/|\vec p|$): 
\begin{equation}\label{poshel}
\vec\sigma\cdot\hat p\;\chi_+=-\chi_+
\;\;. \end{equation} 
It thus describes the negative helicity (left-handed) neutrino, $\nu_L$. Conversely, 
for the negative energy solution with $p^\mu_-=(E<0,-\vec p)$, we obtain:
\begin{equation}\label{neghel}
\vec\sigma\cdot (-\hat p)\;\chi_-=+\chi_-
\;\;, \end{equation} 
thus describing a negative four-momentum positive helicity (right-handed) neutrino, which is 
the positive four-momentum right-handed antineutrino, $\bar\nu_R$.     
  
We conclude that in the Weyl representation the physical $\nu_L\bar\nu_R$-spinor $\psi^{(\nu )}$ is represented by: 
\begin{equation}\label{nuWspinor}
\psi^{(\nu )}_W =\left (\begin{array}{c}\chi =a_+\chi_++a_-\chi_- \\
                                   \phi =0 \end{array}\right )
\;\;, \end{equation} 
where $\chi$ is written as a suitably normalized superposition of the two-spinors $\chi_\pm$, 
e.g., considering plane wave states \cite{HalzenMartin}.   
Next, we calculate 
the corresponding four-spinor in the Dirac-Pauli representation: 
\begin{equation}\label{nuDspinor}
\psi^{(\nu )}_D =U_W\psi^{(\nu )}_W =\frac{1}{\surd 2}\left (\begin{array}{c} \chi \\
                                                 -\chi \end{array}\right )
\;\;, \end{equation}
where we applied the unitary transformation $U$ defined in Eqs.\,(\ref{Utransf}).   

We proceed to calculate the bilinear covariants which are needed in 
Eqs.\,(\ref{spindecomp})-(\ref{antisymt}).   
Using the explicit form of the $\gamma$-matrices in the Dirac-Pauli representation, 
Eqs.\,(\ref{gammaD}), we obtain: 
\begin{equation}\label{zerodensities}
\bar\psi^{(\nu )}\psi^{(\nu )} =0=\bar\psi^{(\nu )}\gamma^5\psi^{(\nu )}  
\;\;, \end{equation} 
with $\bar\psi^{(\nu )}\equiv\psi^{(\nu )\dag}\gamma^0=(\chi^\dag\;\chi^\dag )/\surd 2$\,,  
dropping the subscript $D$ from now on. 
Furthermore: 
\begin{eqnarray}\label{dens0}
&\;&\bar\psi^{(\nu )}\gamma^0\psi^{(\nu )} =\chi^\dag\chi =\bar\psi^{(\nu )}\gamma^5\gamma^0\psi^{(\nu )}
\;\;, \\ [1ex] \label{densi}
&\;&\bar\psi^{(\nu )}\gamma^i\psi^{(\nu )}=-\chi^\dag\sigma^i\chi =\bar\psi^{(\nu )}\gamma^5\gamma^i\psi^{(\nu )}
\;\;. \end{eqnarray}
Finally, using $\sigma^{\mu\nu}\equiv i[\gamma^\mu ,\gamma^\nu ]/2$\,, we also obtain: 
\begin{equation}\label{dens01} 
\bar\psi^{(\nu )}\sigma^{\mu\nu}\psi^{(\nu )}=0
\;\;. \end{equation}
  
Summarizing, only the vector and axial vector (two-point) densities are nonzero for the $\nu_L\bar\nu_R$-system 
and, in fact, they are equal. This yields for the neutrino Wigner function components: 
${\cal V}^{(\nu )}_\mu (x,p)={\cal A}^{(\nu )}_\mu (x,p)$, while all other components vanish, see 
Eqs.\,(\ref{spindecomp})-(\ref{antisymt}). 

Presently we made use only of the algebraic properties of the 
neutrino spinors. It did not enter that the amplitudes $a_\pm$ from Eq.\,(\ref{nuWspinor})  
actually are to be considered as creation/annihilation operators.
  

\end{document}